\documentclass{egpubl}

\JournalSubmission    
 \electronicVersion 

\ifpdf \usepackage[pdftex]{graphicx} \pdfcompresslevel=9
\else \usepackage[dvips]{graphicx} \fi

\PrintedOrElectronic

\usepackage{t1enc,dfadobe}
\usepackage{multirow}
\usepackage{rotating}
\usepackage{egweblnk}
\usepackage{cite}

\let\bf=\bfseries
\let\vec=\mathbf

\let\set=\mathcal
\newcommand{\para}[1]{\noindent{\bf #1}}

\usepackage{amsfonts,amsmath,amssymb}

\usepackage[usenames,dvipsnames]{color}

\usepackage{verbatim}

\newcommand{\bc}{\mathbf{c}}

\newcommand{\bx}{\mathbf{x}}
\newcommand{\by}{\mathbf{y}}

\title[Data-Driven Shape Analysis and Processing]%
      {Data-Driven Shape Analysis and Processing}

\author[K. Xu \& V. Kim \& Q. Huang \& E. Kalogerakis]
       {Kai Xu\thanks{Corresponding author: kevin.kai.xu@gmail.com}$^{1,2}$
        \quad Vladimir G. Kim$^{3}$
        \quad Qixing Huang$^{4}$
        \quad Evangelos Kalogerakis$^{5}$
        \\
         $^1$ Shenzhen VisuCA Key Lab / SIAT \quad
         $^2$ National University of Defense Technology\\
         $^3$ Stanford University \quad
         $^4$ Toyota Technological Institute at Chicago \quad
         $^5$ University of Massachusetts Amherst
       }

\begin{document}

\maketitle

\begin{abstract}

Data-driven methods play an increasingly important role in discovering geometric, structural, and semantic relationships between 3D shapes in collections, and
applying this analysis to support intelligent modeling, editing, and visualization of geometric data.
In contrast to traditional approaches, a key feature of data-driven approaches is that they
aggregate information from a collection of shapes to improve the analysis and processing of individual shapes.
In addition, they are able to learn models that reason about properties and relationships of shapes without relying on hard-coded rules or explicitly programmed instructions. We provide an overview of the main concepts and components of these techniques, and discuss their application to shape classification, segmentation, matching, reconstruction, modeling and exploration, as well as scene analysis and synthesis, through reviewing the literature and relating the existing works with both qualitative and numerical comparisons.
We conclude our report with ideas that can inspire future research in data-driven shape analysis and processing. 

\begin{classification} 
\CCScat{Computer Graphics}{I.3.3}{Picture/Image Generation}{Line and curve generation}
\end{classification}

\end{abstract}


\section{Introduction}
\label{sec:intro}



As big geometric data is becoming more available (e.g., from fast and commodity 3D sensing and crowdsourcing shape modeling), the interest in processing of 3D shapes and scenes has been shifting towards data-driven techniques. These techniques leverage data to facilitate high-level shape understanding, and use this analysis to build effective tools for modeling, editing, and visualizing geometric data. In general, these methods start by discovering patterns in geometry and structure of shapes, and then relate them to high-level concepts, semantics, function, and models that explain those patterns. The learned patterns serve as strong priors in various geometry processing applications.
In contrast to traditional approaches, data-driven methods analyze a set of shapes jointly to extract and model meaningful mappings and correlations in the data,
and learn priors directly from the data instead of relying on hard-coded rules or explicitly programmed instructions.

The idea of utilizing data to support geometry processing has been exploited and practiced for many years. However, most existing works based on this idea are confined to example-based paradigm, thus mostly leveraging only one core concept of data-driven techniques -- \emph{information transfer}. Typically, the input to these problems includes one or multiple exemplar shapes with prescribed or precomputed information of interest, and a target shape that needs to be analyzed or processed. These techniques usually establish a \emph{correlation} between the source and the target shapes and transfer the interesting information from the source to the target. The applications of such approach include a variety of methods in shape analysis (e.g.~\cite{Schaefer:2007:EBS}) and shape synthesis (e.g.~\cite{Merrell:2007:EBM,Ma:2014:ADS}).

As the number of available 3D shapes becomes significantly large, geometry processing techniques supported by these data go through a fundamental change.
Several new concepts emerge in addition to information transfer, opening space for developing new techniques for shape analysis and content creation. In particular, the rich variability of 3D content in existing shape repositories makes it possible to directly reuse the shapes or parts for constructing new 3D models~\cite{Funkhouser:2004:MBE}. \emph{Content reuse} for 3D modeling is perhaps the most straightforward application of big 3D geometric data, providing a promising approach to address the challenging 3D content creation problem. In addition, high-level understanding of shapes can benefit from co-analyzing collections of shapes. Several analysis tools demonstrate that shape analysis is more reliable if it is supported by observing certain attributes in a set of semantically related shapes instead of a single object. \emph{Co-analysis} requires a critical step of finding the correlation between multiple shapes in the input set, which is substantially different from building pair-wise correlation. A key concept to co-analysis is \emph{consistency} of the correlations in the entire set, which has both semantic~\cite{Kalogerakis:2010:LMS,Sidi:2011:CS,Wang:2012:ACS} and mathematical~\cite{Huang:2013:SDP} justifications.

\begin{figure}[t!]
\centering
    \includegraphics[width=0.99\linewidth]{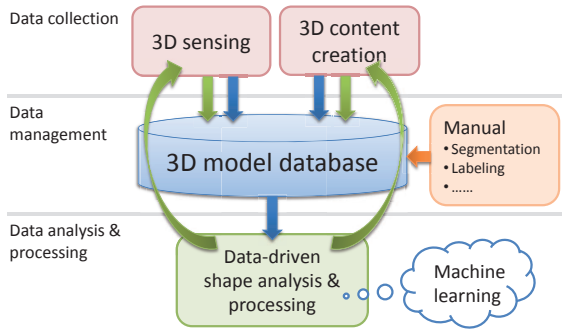}
    \caption{Data-driven shape processing and modeling provides a promising solution to the development of ``big 3D data''.
    Two major ways of 3D data generation, 3D sensing and 3D content creation, populate 3D databases with fast growing amount of 3D models.
    The database models are sparsely enhanced with manual segmentation and labeling, as well as reasonably
    organized, to support data-driven shape analysis and processing, based on, e.g., machine learning techniques.
    The learned knowledge can in turn support efficient 3D reconstruction and 3D content creation, during which
    the knowledge can be transferred to the newly generated data.
    Such 3D data with semantic information can be included into the database to enrich it and facilitate further data-driven applications.}
    \label{fig:teaser}
\end{figure}

\paragraph*{Relation to knowledge-driven shape processing.}
Prior to the emergence of data-driven techniques, high-level shape understanding and modeling was usually achieved with knowledge-driven methods.
In knowledge-driven paradigm, geometric and structural patterns are extracted and interpreted with the help of explicit rules or hand-crafted parameters.
Such examples include heuristics-based shape segmentation~\cite{Shamir:2008:SMS} and procedural shape modeling~\cite{Muller:2006:PMB}.
Although these approaches find certain empirical success, they exhibit several inherent limitations. First, it is extremely hard to hard-code explicit rules and heuristics that can handle the enormous geometric and structural variability of 3D shapes and scenes in general. As a result, knowledge-driven techniques are unlikely to generalize successfully to large and diverse shape collections. Another issue is that it is usually hard for non-expert users to interact with knowledge-driven techniques that require as input ``low-level'' geometric parameters or instructions.

In contrast to knowledge drive methods, data-driven techniques learn representation and parameters from data. Their usually do not depend on hard-coded prior knowledge, and consequently do not rely on hand-crafted parameters, making these techniques more data-adaptive and thus lead to significantly improved performance in many practical settings.
The success of data-driven approaches, backed by machine learning techniques, heavily relies on the accessibility of large data collections.
We have witnessed the success of increasing the training set by orders of magnitude to significantly improve the performance of common machine learning algorithms~\cite{Banko:2001:MPA}. Thus, the recent developments in 3D modeling tools and acquisition techniques for 3D geometry, as well as availability of large repositories of 3D shapes (e.g., Trimble 3D Warehouse, Yobi3D , etc.), offer great opportunities for developing data-driven approaches for 3D shape analysis and processing.

\paragraph*{Relation to structure-aware shape processing.}
This report is closely related to the recent survey on ``structure-aware shape processing'' by Mitra and co-workers~\cite{Mitra:2014:SASP},
which concentrates on techniques for structural analysis of 3D shapes, as well as high-level shape processing guided by structure-preservation.
In that survey, shape structure is defined as the arrangement and relations between shape parts, which is analyzed through identifying shape parts, part parameters, and part relations. Each of the three can be determined through manual assignment, predefined model fitting and data-driven learning.

In contrast, our report takes from a very different perspective---how the availability of big geometric data has changed the field of shape analysis and processing. In particular, we want to highlight several key distinctions:
\emph{First}, data-driven shape processing goes beyond structure analysis.
For example, leveraging large shape collections may benefit a wider variety of problems in shape understanding and processing, such as parametric modeling of shape space~\cite{Allen:2003:SHB}, hypothesis generation for object and scene understanding~\cite{Zia:2013:DR,Satkin:2012:DDS}, and information transfer between multi-modal data~\cite{Wang:2013:PAS,Su:2014:EID}. Data-driven shape processing may also exploit the data-centered techniques in machine learning such as sparse representation~\cite{Ren:2013:HSC} and feature learning~\cite{Lai:2013:UFL}, which are not pre-conditioned on any domain-specific or structural prior beyond raw data.
\emph{Second}, even within the realm of structure-aware shape processing, data-driven approaches are arguably becoming the dominant branch due to their theoretical and practical advantages, availability of large shape repositories, and recent developments in machine learning.

\begin{figure*}[t!]
\centering
    \includegraphics[width=0.9\textwidth]{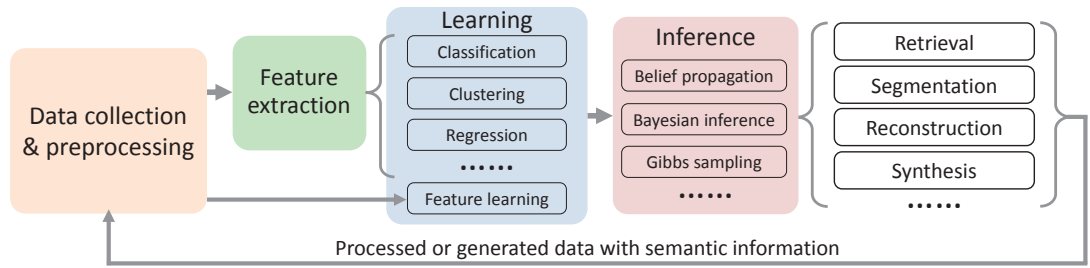}
    \caption{
The general pipeline of data-driven geometry processing contains four major stages: data collection and preprocessing, feature extraction (or feature learning),
learning and inference. The inference supports many applications which would produce new shapes or scenes through reconstruction modeling or synthesis.
These new data, typically possessing labels for shapes or parts, can be used to enrich the input datasets and enhance the learning tasks in future, forming a data-driven geometry processing loop.}
    \vspace{-.6cm}
    \label{fig:overview}
\end{figure*}

\paragraph*{Vision and motivation.}
With the emergence of ``big data'', many scientific disciplines have shifted their focus to data-driven techniques. Although 3D geometry data is still far from being as ubiquitous as some other data formats (e.g., photographs), rapidly growing number of 3D models, recent developments in fusing 2D and 3D data, and invention of commodity depth cameras,
have made the era of ``big 3D data'' more promising than ever.
%
At the same time, we expect data-driven approaches to take one of the leading roles in understanding and reconstruction of acquired 3D data, as well as synthesis of new shapes.
In summary, data-driven geometry processing will close the loop from acquisition, analysis, and processing to generation of 3D shapes (see Figure~\ref{fig:teaser}), and will be a key tool for manipulating big visual data.

Recent years have witnessed a rapid development of data-driven geometry processing algorithms, both in computer graphics and in computer vision communities. Given the research efforts and wide interests in the subject, we believe many researchers would benefit from a comprehensive and systematic survey. We also wish such a survey can simulate new theories, problems, and applications,

\paragraph*{Organization.}
This survey is organized as follows. Section~\ref{sec:overview} gives a high-level overview of data-driven approaches and classifies data-driven methods with respect to their application domains. This section also provides two representative examples for the reader to understand the general work-flow of data-driven geometry processing. The following sections survey various data-driven shape processing problems in detail. Finally, we conclude by listing a set of key challenges and providing a vision on future directions.

\paragraph*{Accompanying online resources.}
In order to assist the readers in learning and leveraging the basic algorithms, we provide an online wikipage~\cite{Wikipage}, which collects tools, source codes, together with benchmark data for typical problems and applications of data-driven shape processing. This page will also provide links and data mining tools for obtaining large data collections of shapes and scenes. The website would serve as a starting point for those who are conducting research in this direction, we also expect it to benefit a wide spectrum of researchers from related fields.

\section{Overview}
\label{sec:overview}

In this section, we provide a high-level overview of the main components and steps of data-driven approaches for processing 3D shapes and scenes. Although the pipeline of these methods significantly vary depending on their particular applications and goals, a number of components tend to be common: the input data collection and processing, data representations and feature extraction, learning and inference. \emph{Representation, learning and inference} are critical components of machine learning approaches in general \cite{Koller:2009:PGM}. In the case of shape and scene processing, each of these components poses several interesting and unique problems when dealing with 3D geometric data. These problems have greatly motivated the research on data-driven geometry processing, and in turn brought new challenges to computer vision and machine learning communities, as reflected by the increasing interest in 3D visual data from these fields. Below, we discuss particular characteristics and challenges of data-driven 3D shape and scene processing algorithms.
Figure~\ref{fig:overview} provides a schematic overview of the most common components of these algorithms.

\begin{figure*}[t!]
\centering
    \includegraphics[width=1.0\textwidth]{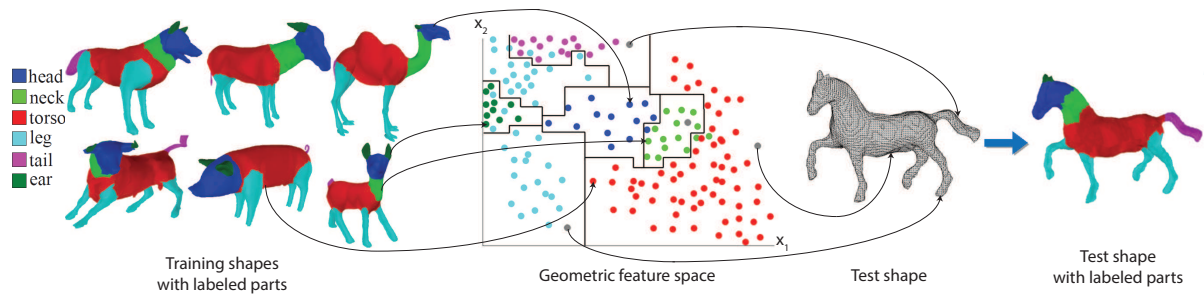}
    \vspace{-.5cm}
    \caption{
Pipeline of a supervised segmentation algorithm \cite{Kalogerakis:2010:LMS}. Given a set of shapes with labeled parts, the points of each shape are embedded in a common feature space based on their local geometric descriptors (a color is assigned to points depending on their given part label). A classifier is learned to split the feature space into regions corresponding to each part label. Given a test shape, its points (shown in grey) are first embedded in the same space. Then part labels are inferred for all its points based on the learned classifier and an underlying structured probabilistic model (Section \ref{sec:segmentation}).}
    \vspace{-.6cm}
    \label{fig:overview_seg}
\end{figure*}

\subsection{3D data collection}

\para{Shape representation.} A main component of data-driven approaches for shape and scene processing is data collection, where the goal is acquire a number of 3D shapes and scenes depending on the application. When shapes and scenes are captured with scanners or depth sensors, their initial representation is in the form of \emph{range data} or \emph{unorganized point clouds}. Several data-driven methods for reconstruction, segmentation and recognition directly work on these representations and do not require any further processing. On the other hand, online repositories, such as the Trimble 3D Warehouse, contain millions of shapes and scenes that are represented as \emph{polygon meshes}. A large number of data-driven techniques are designed to handle complete shapes in the form of polygon meshes created by 3D modeling tools or re-constructed from point clouds. Choosing which representation to use depends on the application. For example, data-driven reconstruction techniques aim for generating complete shapes and scenes from noisy point clouds with missing data. The reconstructed shapes can then be processed with other data-driven methods for categorization, segmentation, matching and so on. Developing methods that can handle any 3D data representation, as well as jointly reconstruct and analyze shapes is a potential direction for future research we discuss in Section \ref{sec:conclusion}.

When polygon meshes are used as the input representation, an important aspect to consider is whether and how data-driven methods will deal with possible ``defects'', such as non-manifold and non-orientable sets of polygons, inverted faces, isolated elements, self-intersections, holes and topological noise. The vast majority of meshes available in online repositories have these problems. Although there is a number of mesh repairing tools (see \cite{Campen:PGP:2012} for a survey), they may not handle all different types of ``defects'', and can take a significant amount of time to process each shape in large datasets. To avoid the issues caused by these ``defects'', some data-driven methods uniformly sample the input meshes and work on the resulting point-based representation instead (e.g., \cite{Chaudhuri:2011:prabm,Kim:2013:lpt}).

\paragraph*{Datasets.} Although it is desirable to develop data-driven methods that can learn from a handful of training shapes or scenes, this is generally a challenging problem in machine learning \cite{Fei:OSL:2006}. Several data-driven methods in computer vision have been particularly successful due to the use of very large datasets that can reach the size of several millions of images \cite{Torralba:2008:MTI}. In contrast, data-driven approaches for 3D shape and scene processing approaches have mostly relied on datasets that reach the order of a few thousands so far (e.g., Princeton Shape Benchmark \cite{Shilane:2004:TPS}, or datasets collected from the web \cite{Kim:2013:lpt}). Online repositories contain large amount of shapes, which can lead to the development of methods that will leverage datasets that are orders of magnitudes larger than the ones currently used. Another possibility is to develop synthetic datasets. A notable example is the pose and part recognition algorithm used in Microsoft's Kinect that relies on 500K synthesized shapes of human bodies in different poses \cite{Shotton:2011:RLH}. In general, large datasets are important to capture the enormous 3D shape and scene variability, and can significantly increase the predictive performance and usability of learning methods. A more comprehensive summary of the existing online data collections can be found on our wikipage~\cite{Wikipage}.

\subsection{3D data processing and feature representation}
It is common to perform some additional processing on the input representations of shapes and scenes before executing the main learning step. The reason is that the input representations of 3D shapes and scenes can have different resolutions (e.g., number of points or faces), scale, orientation, and structure. In other words, the input shapes and scenes do not initially have any type of common parameterization or alignment. This is significantly different from other domains, such as natural language processing or vision, where text or image datasets frequently come with a common parameterization beforehand (e.g., images with the same number of pixels and objects of consistent orientation).

To achieve a common parameterization of the input shapes and scenes, one popular approach is to embed them in a \emph{common geometric feature space}. For this purpose a variety of shape descriptors have been developed. These descriptors can be classified into two main categories: \emph{global shape descriptors} that convert each shape to a feature vector,  and \emph{local shape descriptors} that convert each point to a feature vector. Examples of global shape descriptors are Extended Gaussian Images \cite{Horn:1984:EGI}, 3D shape histograms \cite{Ankerst:1999:3dsh,Chaudhuri:2010:ddsc}, spherical functions \cite{Saupe:2001:MRS}, lightfield descriptors \cite{Chen:2003:ovsb}, shape distributions \cite{Osada:2002:sd}, symmetry descriptors \cite{Kazhdan:2004:SD3D}, spherical harmonics \cite{Kazhdan:2003:RISH}, 3D Zernicke moments \cite{Novotni:2003:3DZD}, and bags-of-words created out of local descriptors \cite{Bronstein:2011:SGGW}. Local shape descriptors include surface curvature, PCA descriptors, local shape diameter \cite{Shapira:2008:SDF}, shape contexts \cite{Belongie:2002:SMO,Kalogerakis:2010:LMS,Kokkinos:2012:ISC}, spin images \cite{Johnson:1999:USI}, geodesic distance features \cite{Zhang:2005:FSP}, heat-kernel descriptors \cite{Bronstein:2011:SGGW}, and depth features \cite{Shotton:2011:RLH}. Global shape descriptors are particularly useful for shape classification, retrieval and organization. Local shape descriptors are useful for partial shape matching, segmentation, and point correspondence estimation. Before using any type of global or local descriptor, it is important to consider whether the descriptor will be invariant to different shape orientations, scales, or poses. In the presence of noise and irregular mesh tessellations, it is important to robustly estimate local descriptors, since surface derivatives are particularly susceptible to surface and sampling noise \cite{Kalogerakis:2007:RSE}.

Sometimes it is common to use several different descriptors, and let the learning step decide which ones are more relevant for each class of shapes \cite{Kalogerakis:2010:LMS}. A promising future direction is to develop data-driven methods that learn feature representations from raw 3D geometric data, enlightened by the recent hot topic of deep learning~\cite{Bengio:2009:LDA}. Similar direction is already explored in computer vision for 2D images \cite{Yu:FLI:2010}. In 3D, some works attempt feature learning on the volumetric representation of 3D shapes or essentially 3D images~\cite{Lai:2013:UFL}. A more popular approach is to apply deep learning directly on the raw RGB-D data captured by a depth camera~\cite{Socher:2012:CRD,Blum:2012:LFD,Bo:2014:LHS}.

Instead of embedding shapes in a common geometric feature space, several methods instead try to directly align shapes in Euclidean space. We refer the reader to the survey on dynamic geometry processing for a tutorial on rigid and non-rigid registration techniques \cite{Chang:DGP:2012}. An interesting extension of these techniques is to include the alignment process in the learning step of data-driven methods, since it is inter-dependent with other shape analysis tasks such as shape segmentation and correspondences \cite{Kim:2013:lpt}.

Some data-driven methods require additional processing steps on the input. For example, learning deformation handles or fully generative models of shapes usually rely on segmenting the input shapes into parts with automatic algorithms \cite{Huang:2011:JSS,Sidi:2011:CS} and representing these parts with surface abstractions \cite{Yumer:2012:CSC} or descriptors \cite{Kalogerakis:2012:PMC}. To decrease the amount of computation required during learning, it is also common to represent the shapes as a set of patches (super-faces) \cite{Huang:2011:JSS} inspired by the computation of super-pixels in image segmentation.

\subsection{Learning and Inference}
The processed representations of shapes and scenes are used to perform learning and inference for a variety of applications: shape classification, segmentation, matching, reconstruction, modeling, synthesis, scene analysis and synthesis. The learning procedures significantly vary depending on the application, thus we discuss them individually in each of the following sections on these applications. As a common theme, learning is viewed as an \emph{optimization} problem that runs on a set of variables representing geometric, structural, semantic or functional properties of shapes and scenes. There is usually a single or multiple objective (or loss) functions for quantifying preferences for different models or patterns governing the 3D data. After learning a model from the training data, inference procedures are used to predict values of variables for new shapes or scenes. Again, the inference procedures vary depending on the application, and are discussed separately in the following sections. It is common that inference itself is an optimization problem, and sometimes is part of the learning process when there are latent variables or partially observed input shape or scene data.

A general classification of the different types of algorithms used in data-driven approaches for shape and scene processing can be derived from the type of input information available during learning:

\begin{itemize}
\item \textbf{Supervised learning} algorithms are trained on a set of shapes or scenes annotated with labeled data. For example, in the case of shape classification, these labeled data can have the form of tags, while in the case of segmentation, the labeled data have the form of segmentation boundaries or part labels. The labeled data can be provided by humans or generated synthetically. After learning, the learned models are applied on different sets of shapes (test shapes) to produce results relevant to the task.
\item \textbf{Unsupervised} algorithms co-analyze the input shapes or scenes without any additional labeled data i.e., the desired output is unknown beforehand. The goal of these methods is to discover correlations in the geometry and structure of the input shape or scene data. For example, unsupervised shape segmentation methods usually perform some type of clustering in the feature space of points or patches belonging to the input shapes.
\item \textbf{Semi-supervised} algorithms make use of shapes (or scenes) with and without any labeled data. Active learning is a special case of semi-supervised learning in which a learning algorithm interactively queries the user to obtain desired outputs for more data points related to shapes.
\end{itemize}

In general, supervised methods tend to output results that are closer to what a human would expect given the provided labeled data, however they may fail to produce desirable results when the training shapes (or scenes) are largely geometrically and structurally dissimilar with the test shapes (or scenes). They also tend to require a substantial amount of labeled information as input, which can become a significant burden for the user. Unsupervised methods can deal with collections of shapes and scenes with larger variability and require no human supervision. However, they sometimes require parameter tuning to yield the desired results. Semi-supervised methods represent a trade-off between supervised and unsupervised methods: they provide more direct control to the user about the desired result compared to unsupervised methods, and often produce considerable improvements in the results by making use of both labeled and unlabeled shapes or scenes compared to supervised methods.

\begin{figure}[t]
\centering
    \includegraphics[width=1.0\columnwidth]{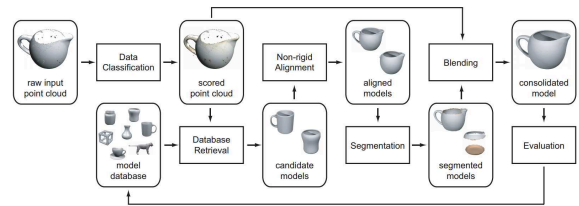}
    \vspace{-0.5cm}
    \caption{The data-driven shape reconstruction pipeline proposed in~\cite{Pauly:2005:ESC}.}
    \vspace{-0.6cm}
    \label{fig:pauly_sgp05_esr}
\end{figure}

\paragraph*{The data-driven loop.}
An advantageous feature of data-driven shape processing is that the output data, produced by learning and inference, typically come with rich semantic information.
For example, data-driven shape segmentation produces parts with semantic labels~\cite{Kalogerakis:2010:LMS}; data-driven reconstruction is commonly coupled with semantic part or shape recognition~\cite{Shen:2012:SRP,Nan:2012:SAC}; data-driven shape modeling can generate readily usable shapes inheriting the semantic information from the input data~\cite{Xu:2011:PMO}.
These processed and generated data can be used to enrich the existing shape collections with both training labels and reusable contents, which in turn benefit subsequent learning. In a sense,
data-driven methods \emph{close the loop of data generation and data analysis} for 3D shapes and scenes; see Figure~\ref{fig:overview}.
Such concept has been practiced in several prior works, such as the data-driven shape reconstruction framework proposed in~\cite{Pauly:2005:ESC} (Figure~\ref{fig:pauly_sgp05_esr}).

\paragraph*{Pipeline example.}
To help the reader grasp the pipeline of data-driven methods, a schematic overview of the components in Figure \ref{fig:overview}.
Depending on the particular application, the pipeline can have several variations, or some components might be skipped. We discuss the main components and steps of algorithms for each application in more detail in the following sections. A didactic example of the pipeline in the case of supervised shape segmentation is shown in Figure \ref{fig:overview_seg}. The input shapes are annotated with labeled part information. A geometric descriptor is extracted for each point on the training shapes, and the points are embedded in a common feature space. The learning step uses a classification algorithm that non-linearly separates the input space into a set of regions corresponding to part labels in order to optimize classification performance (more details are provided in Section \ref{sec:segmentation}). Given a test shape, a probabilistic model is used to infer part labels for each point on that shape based on its geometric descriptor in the feature space.

\subsection{A comparative overview}
Before reviewing the related works in detail under various applications, we provide a comparative overview of the entire body of works
to be reviewed in this survey (see Table~\ref{tab:compare}), to correlate these methods under a set of \emph{criteria} for data-driven approach to shape analysis and processing:
\begin{itemize}
  \item \textbf{Training data.} We concern about the representation, pre-processing and scale of \emph{training} data. Note that once a model is learned from the training data,
                                it can be used to inference on test data of different modality.
                                For single shapes, the mostly adopted representations are mesh model and point cloud. 3D scenes are typically represented as an arrangement of
                                individual objects (mesh model).
                                Pre-processing include pre-segmentation, over-segmentation, pre-alignment, initial correspondence, and labeling.
  \item \textbf{Feature.} Roughly speaking, there are two types of features involved in data-driven shape processing.
                          The most commonly used features are low-level ones, such as local geometric features (e.g., local curvature)
                          and global shape descriptor (e.g. shape distribution~\cite{Osada:2002:sd}). If the input shapes are pre-segmented into meaningful parts, high-level structural
                          features (spatial relationship) can be derived. Generally, working with high-level features enables the learning of more powerful models
                          for more advanced inference tasks, such as structural analysis~\cite{Mitra:2014:SASP}, on more complex data such as man-made objects and scenes.
  \item \textbf{Learning model/approach.} The specific choice of learning method is application-dependent. In most cases, machine learning approaches are adapted to
                                          geometric data with feature extraction. For some problems, such as shape correspondence, the core problem is to extract geometric correlation
                                          between different shapes, in an unsupervised manner, which itself can be seen as a learning problem specific to geometry processing.
  \item \textbf{Learning type.} As discussed above, there are three basic types of data-driven approaches, depending on the availability of labeled training data:
                                supervised, semi-supervised and unsupervised.
  \item \textbf{Learning outcome.} The learning would produce a parametric or non-parametric model (classifier, clustering, regressor, etc.) used for inference,
                                   a learned distance metric which can be utilized for further analysis, and/or feature representations learned from raw data.
  \item \textbf{Application.} The main applications of data-driven shape analysis and processing are: classification, segmentation, correspondence,
                              modeling, synthesis, reconstruction, exploration and organization.
\end{itemize}

\section{Shape Classification}
\label{sec:classification}
Data-driven techniques commonly make assumptions about the size and homogeneity of the input data set. In particular, existing analysis techniques often assume that all models belong to the same class of objects~\cite{Kim:2013:lpt} or scenes~\cite{Fisher:2011:CSR}, and cannot directly scale to entire repository such as the Trimble 3D Warehouse~\cite{warehouse}. Similarly, techniques for data-driven reconstruction of indoor environments assume that the input data set only has furniture models~\cite{Nan:2012:SAC}, while modeling and synthesis interfaces restrict the input data to particular object or scene classes~\cite{Chaudhuri:2011:prabm,Kalogerakis:2012:PMC,Fisher:2012:CSR}.  Thus, as a first step these methods query a 3D model repository to retrieve a subset of relevant models.

Most public shape repositories such as 3D Warehouse~\cite{warehouse} rely on the users to provide tags and names of the shapes with little additional quality control measures. As a result, the shapes are sparsely labeled with inconsistent and noisy tags.  This motivates developing automatic algorithms to infer text associated with models.  Existing work focuses on establishing class memberships for an entire shape (e.g. this shape is a chair), as well as inferring finer-scale attributes (e.g. this chair has a rocking leg).

\paragraph*{Classification} methods assign a class membership for unlabeled shapes. One approach is to retrieve for each unlabeled shape the most similar shape from a database of 3D models with known shape classes. There has been a large number of shape descriptors proposed in recent years that can be used in such a retrieval task, and one can refer to the survey of Tangelder et al. \cite{Tangelder:2008:ASC} for a thorough overview.
One can further improve classification results by leveraging machine learning techniques to learn classifiers that are based on global shape descriptors~\cite{Frome:2004:rord,Golovinskiy:2009:SBR3D}. Barutcuoglu et al.~\cite{Barutcuoglu:2006:hscu} demonstrate that Bayesian aggregation can be used to improve classification of shapes that are a part of a hierarchical ontology of objects. Bronstein et al.\cite{Bronstein:2011:SGGW} leverage ``bag of features'' to learn powerful descriptor-space metrics for non-rigid shapes. These technique can be further improved by using sparse coding techniques~\cite{Litman:2014:SLBF}.

\begin{figure}[bt]
\centering
    \includegraphics[width=1.0\columnwidth]{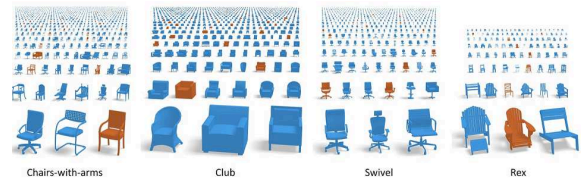}
   \vspace{-.6cm}
    \caption{
    Fine-grained classification of 3D models \cite{Huang:2013:FSL}, where text labels are propagated from brown to blue models. }
   \vspace{-.4cm}
    \label{fig:fine-grained}
\end{figure}

\paragraph*{Tag attributes} often capture fine-scale attributes of shapes that belong to the same class. These attributes can include presence or absence of particular parts, object style, or comparative adjectives. Huang et al.~\cite{Huang:2013:FSL} developed a framework for propagating these attributes in a collection of partially annotated 3D models. For example, only brown models in Figure \ref{fig:fine-grained} were labeled, and blue models were annotated automatically. To achieve automatic labeling, they start by co-aligning all models to a canonical domain, and generate a voxel grid around the co-aligned models. For each voxel they compute local shape features, such as spin images, for each shape. Then, they learn a distance metric that best discriminates between different tags. All shapes are finally embedded in a weighted feature space where nearest neighbors are connected in a graph. A graph cut clustering is used to assign tags to unlabeled shapes.

\begin{figure}[tb]
\centering
    \includegraphics[width=1.0\columnwidth]{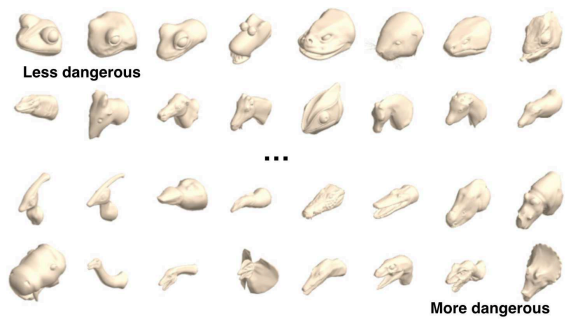}
    \vspace{-.6cm}
    \caption{
    Ranking of parts with respect to ``dangerous'' attribute (image from \cite{Chaudhuri:2013:ACC}) }
    \vspace{-.6cm}
    \label{fig:attribit}
\end{figure}

While above method works well for discrete tags, it does not capture more continuous relations, such as animal A is more dangerous than animal B.  Chaudhuri et al.~\cite{Chaudhuri:2013:ACC} focus on estimating ranking based on comparative adjectives. They ask people to compare pairs of shape parts with respect to different adjectives, and use a Support Vector Machine ranking method to predict attribute strengths from shape features for novel shapes (Figure \ref{fig:attribit}).

While the techniques described above are suitable for retrieving related models, most of the described method are not designed to understand intra-class variations. Usually a more involved structural analysis is necessary to understand higher-level semantic properties of shapes. Even for inferring tag attributes existing works relies on shape matching~\cite{Huang:2013:FSL} or shape segmentation~\cite{Chaudhuri:2013:ACC}. The following two sections will focus on inferring these higher-level structural properties in collections of shapes.


\begin{table*}[t!]
\small
  \centering
    \begin{tabular*}{\textwidth}{l|c|c|c|c|c@{}}
    \hline
    Segmentation                & Learning           & Type of          & PSB rand index (\# train. & L-PSB  accuracy (\# train.  & COSEG          \\
    method                      & type               & manual input     & shapes if applicable)     & shapes if applicable)       & accuracy       \\
    \hline
    \hline
    \cite{Kalogerakis:2010:LMS} & supervised         & labeled shapes   & 9.4\% (19) / 14.8\% (3)   & 95.3\% (19) / 89.2\% (3)    & unknown        \\
    \hline
    \cite{Benhabiles:2011:LBE}  & supervised         & segmented shapes & 8.8\% (19) / 9.7\% (6)    & not applicable              & not applicable \\
    \hline
    \cite{Huang:2011:JSS}       & unsupervised       & none             & 10.1\%                    & not applicable              & not applicable \\
    \hline
    \cite{Sidi:2011:CS}         & unsupervised       & none             & unknown                   & unknown                     & 87.7\%         \\
    \hline
    \cite{van-Kaick:2011:PKC}   & supervised         & labeled shapes   & unknown                   & \~88.7\% (12), see caption   & unknown        \\
    \hline
    \cite{Hu:2012:CSS}          & unsupervised       & none             & unknown                   & 88.5\%                      & 91.4\%         \\
    \hline
    \cite{Lv:2012:SMS}          & semi-supervised    & labeled shapes   & unknown                   & 92.3\% (3)                  & unknown        \\
    \hline
    \cite{Wang:2012:ACS}        & semi-supervised    & link constraints & unknown                   & unknown                     & `close to error-free' \\
    \hline
    \cite{Wang:2013:PAS}        & supervised         & labeled images   & unknown                   & \~88.0\% (19), see caption   & unknown               \\
    \hline
    \cite{Kim:2013:lpt}         & semi-/unsupervised & box templates    & unknown                   & unknown                     & 92.7\% (semi-superv.) \\
    \hline
    \cite{Huang:2014:FMN}       & unsupervised       & none             & unknown                   & unknown                     & 90.1\%                \\
    \hline
    \cite{Xu:2014:TSS}          & supervised         & labeled shapes   & 10.0\%                    & 86.0\%                      & unknown               \\
    \hline
    \cite{Zhige:2014:SSL}       & supervised         & labeled shapes   & 10.2\% (19)               & 94.2 (19) / 88.6 (5)        & unknown               \\
    \hline
    \end{tabular*}%
  \caption{Performance of data-driven methods for segmentation in the Princeton Segmentation Benchmark (PSB) and COSEG datasets.
		   Left to right: segmentation method, learning type depending on the nature of data required as input to the method, type of manual input if such required, segmentation performance expressed by the rand index metric \cite{Chen:2009:BMS}, labeling accuracy \cite{Kalogerakis:2010:LMS} based on the PSB and COSEG datasets. 	
  		   We report the rand index segmentation error metric averaged over all classes of the PSB benchmark.
  		   The labeling accuracy is averaged over the Labeled PSB (L-PSB) benchmark excluding the ``Bust'', ``Mech'', and ``Bearing'' classes. The reason is that there are no clear semantic correspondences between parts in these classes, or the ground-truth segmentations do not sufficiently capture semantic parts in their shapes.
  		   We report the labeling accuracy averaged over the categories of the COSEG dataset used in \cite{Sidi:2011:CS}. The COSEG classes ``iron'', ``large chairs'', ``large vases'', ``tele-aliens'' were added later and are excluded here since most papers frequently do not report performance in those.
  		   We note that van Kaick et al.~\cite{van-Kaick:2011:PKC} reported the labeling accuracy in ten of the L-PSB classes, while Wang et al.~\cite{Wang:2013:PAS}  reported the labeling accuracy in seven of the L-PSB classes.
  		   The method by Kim et al.~\cite{Kim:2013:lpt} can run in either semi-supervised or unsupervised mode. In unsupervised mode, the corresponding labeling accuracy is 89.9\% in the COSEG dataset on average.
  		   }
  \label{tab:segmentation_performance}%
\end{table*}

\section{Data-driven Shape Segmentation}
\label{sec:segmentation}

The goal of data-driven shape segmentation is to partition the shapes of an input collection into parts, and also estimate part correspondences across these shapes. We organize the literature on shape segmentation into the following three categories: supervised segmentation, unsupervised segmentation, and semi-supervised segmentation following the main classification discussed in Section \ref{sec:overview}. Table \ref{tab:segmentation_performance} summarizes representative techniques and reports their segmentation and part labeling performance based on established benchmarks. Table \ref{tab:segmentation_running_times} reports characteristic running times for the same techniques.



\subsection{Supervised shape segmentation}

\begin{figure}[t]
\centering
    \includegraphics[width=1.0\columnwidth]{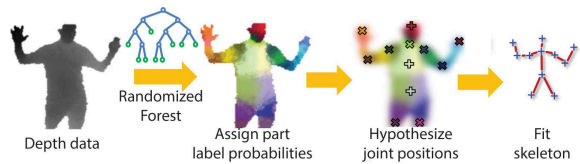}
    \vspace{-0.4cm}
    \caption{
    A random forest classifier applied on depth data representing a human body shape (image from \cite{Fossati:2013:CDC}) }
    \vspace{-0.6cm}
    \label{fig:kinect}
\end{figure}

\paragraph*{Classification techniques.} Supervised shape segmentation is frequently formulated as a classification problem. Given a training set of shapes containing points, faces or patches that are labeled according to a part category (see Figure \ref{fig:overview_seg}), the goal of a classifier is to identify which part category other points, faces, or patches from different shapes belong to. Supervised shape segmentation is executed in two steps: during the first step, the parameters of the classifier are learned from the training data. During the second step, the classifier is applied on new shapes. A simple linear classifier has the form:
\begin{equation}
c = f ( \sum\limits_j \theta_j \cdot x_j )
\end{equation}
where $x_j$ is a geometric feature of a point (face, or patch), such as the ones discussed in Section \ref{sec:overview}. The parameters $\theta_j$ serve as weights for each geometric feature. The function $f$ is non-linear and maps to a discrete value (label), which is a part category, or to probabilities per category. In general, choosing a good set of geometric features that help predicting part labels, and employing classifiers that can discriminate the input data points correctly are important design choices. There is no rule of thumb on which is the best classifier for a problem. This depends on the underlying distribution and characteristics of the input geometric features, their dimensionality, amount of labeled data, existence of noise in the labeled data or shapes, training and test time constraints - for a related discussion on how to choose a classifier for a problem, we refer the reader to \cite{Manning:2008:IIR}. Due to the large dimensionality and complexity of geometric feature spaces, non-linear classifiers are more commonly used. For example, to segment human bodies into parts and recognize poses, the Microsoft's Kinect uses a random forest classifier trained on synthetic depth images of humans of many shapes and sizes in highly varied poses sampled from a large motion capture database \cite{Shotton:2011:RLH} (Figure \ref{fig:kinect}).

\paragraph*{Structured models.} For computer graphics applications, it is important to segment shapes with accurate and smooth boundaries. For example, to help the user create a new shape by re-combining parts from other shapes \cite{Funkhouser:2004:MBE}, irregular and noisy segmentation boundaries can cause problems in the part attachment. From this aspect, using a classifier per point/face independently is usually not enough. Thus, it is more common to formulate the shape segmentation problem as an energy minimization problem that involves a unary term assessing the consistency of each point/face with each part label, as well as a pairwise term assessing the consistency of neighboring points/faces with pairs of labels. For example, pairs of points that have low curvature (i.e., are on flat surface) are more likely to have the same part label. This energy minimization formulation has been used in several single-shape and data-driven segmentations (unsupervised or supervised) \cite{Katz:2003:HMD,Anguelov:2005:DLM,Shapira:2010:CPA,Kalogerakis:2010:LMS}. In the case of supervised segmentation \cite{Kalogerakis:2010:LMS}, the energy can be written as:
\begin{equation}
E(\bc; \theta) = \sum_i E_{unary}(c_i; \bx_i, \theta_1) + \sum_{i,j} E_{pairwise}(c_i, c_j; \by_{ij}, \theta_2)
\label{eqn:CRFEnergy}
\end{equation}
where $\bc=\{c_i\}$ is a vector of random variables representing the part label per point (or face) $i$, $\bx_i$ is its geometric feature vector, $i,j$ are indices to points (or faces) that are considered neighbors, $\by_{ij}$ is a geometric feature vector representing dihedral angle, angle between normals, or other features, and $\theta=\{\theta_1,\theta_2\}$  are the energy parameters.  The important difference of supervised data-driven methods with previous single-shape segmentation methods is that the parameters $\theta$ are automatically learned from the training shapes to capture complex feature space patterns per part \cite{Anguelov:2005:DLM,Kalogerakis:2010:LMS}. We also note that the above energy of Equation \ref{eqn:CRFEnergy}, when written in an exponentiated form and normalized, can be treated as a probabilistic graphical model \cite{Koller:2009:PGM}, called Conditional Random Field \cite{Lafferty:2001:CRF} that represents the joint probability distribution over part labels conditioned on the input features:
\begin{equation}
P(\bc | \bx, \by, \theta)= \exp(-E(\bc;\theta))/ Z(\bx,\by,\theta)
\label{eqn:CRFConditional}
\end{equation}
where $Z(\bx,\by,\theta)$ is a normalization factor, also known as partition function. Minimizing the energy of Equation \ref{eqn:CRFEnergy}, or correspondingly finding the assignment $\bc$ that maximizes the above probability distribution is known as a Maximum A Posteriori inference problem that can be solved in various manners, such as graph cuts, belief propagation, variational or linear programming relaxation techniques \cite{Koller:2009:PGM}.

The parameters $\theta$ can be jointly learned through maximum likelihood (ML) or maximum a posteriori (MAP) estimates \cite{Koller:2009:PGM}. However, due to high computational complexity of ML or MAP learning and the non-linearity of classifiers used in shape segmentation, it is common to train the parameters $\theta_1$ and $\theta_2$ of the model separately i.e., train the classifiers of the unary and pairwise term separately \cite{Sutton:2005:PTU}. The exact form of the unary and pairwise terms vary across supervised shape segmentation methods: the unary term can have the form of a log-linear model \cite{Anguelov:2005:DLM}, cascade of JointBoost classifiers \cite{Kalogerakis:2010:LMS}, Gentleboost \cite{van-Kaick:2011:PKC}, or feedforward neural networks \cite{Zhige:2014:SSL}. The pairwise term can have the form of a learned log-linear model \cite{Anguelov:2005:DLM}, label-dependent GentleBoost classifier \cite{Kalogerakis:2010:LMS}, or a smoothness term based on dihedral angles and edge length tuned by experimentation \cite{Shapira:2010:CPA,van-Kaick:2011:PKC,Zhige:2014:SSL}. Again the form of the unary and pairwise terms depend on the amount of training data, dimensionality and underlying distribution of geometric features used, and computational cost.

\begin{table*}[t!]
\small
  \centering
    \begin{tabular}{l|c|c|c}
    \hline
    Segmentation                & Reported                     &  Dataset size for       & Reported \\
    method                      & running times                &  reported running times & processor \\
    \hline
    \hline
    \cite{Kalogerakis:2010:LMS} & 8h train. / 5 min test.      &  6 train. shapes / 1 test shape     & Intel Xeon E5355 2.66GHz \\
    \hline
    \cite{Benhabiles:2011:LBE}  & 10 min train. / 1 min test.  &  unknown for train. / 1 test shape  & Intel Core 2 Duo 2.99GHz \\
    \hline
    \cite{Huang:2011:JSS}       & 32h                          &  380 shapes                         & unknown, 2.4 GHz \\
    \hline
    \cite{Sidi:2011:CS}         & 10 min                       &  30 shapes                          & AMD Opteron 2.4GHz \\
    \hline
    \cite{van-Kaick:2011:PKC}   & 10h train. / few min test.   &  20-30 train. shapes / 1 test shape & AMD Opteron 1GHz  \\
    \hline
    \cite{Hu:2012:CSS}          & 8 min (excl. feat. extr.)    &  20 shapes                          & Intel dual-core 2.93GHz  \\
    \hline
    \cite{Lv:2012:SMS}          & 7h train. / few min test.    &  20 shapes                          & Intel I7 2600 3.4GHz \\
    \hline
    \cite{Wang:2012:ACS}        & 7 min user interaction       &  28 shapes                          & unknown \\
    \hline
    \cite{Wang:2013:PAS}        & 1.5 min (no train. step)     &  1 test shape                       & unknown \\
    \hline
    \cite{Kim:2013:lpt}         & 11h                          &  7442 shapes                        & unknown \\
    \hline
    \cite{Huang:2014:FMN}       & 33h                          &  8401 shapes                        & unknown, 3.2GHZ \\
    \hline
    \cite{Xu:2014:TSS}          & 30 sec (no train. step)      &  1 test shape                       & Intel I5 CPU \\
    \hline
    \cite{Zhige:2014:SSL}       & 15 sec train. (excl. feat. extr.) &  6 train. shapes                & Intel Quad-Core 3.2 GHz \\
    \hline
    \end{tabular}%
  \caption{Running times reported for the data-driven segmentation methods of Table \ref{tab:segmentation_performance}. We note that running times are reported in different dataset sizes and processors in the referenced papers, while it is frequently not specified whether the execution uses one or multiple threads or whether the running times include all the algorithm steps, such as super-face or feature extraction.
           Exact processor information is also frequently not provided. Thus, the reported running times of this table are only indicative and should not serve as a basis for a fair comparison.}
  \label{tab:segmentation_running_times}%
\end{table*}

\paragraph*{Joint labeling.} Instead of applying the learned probabilistic model to a single shape, an alternative approach is to find correspondences between faces of pairs of shapes, and incorporate a third ``inter-shape'' term in the energy of Equation \ref{eqn:CRFEnergy} \cite{van-Kaick:2011:PKC}. The ``inter-shape'' term favors pairs of corresponding faces on different shapes to have the same label. As a result, the energy can be minimized jointly over a set of shapes to take into account any additional correspondences.

\paragraph*{Boundary learning.} Instead of applying a classifier per mesh point, face or patch to predict a part label, a different approach is to predict the probability of each polygon mesh edge to serve as a segmentation boundary or not \cite{Benhabiles:2011:LBE}. The problem can be formulated as a binary classifier (e.g., Adaboost) that is trained from human segmentation boundaries. The input to the classifier are geometric features of edges, such as dihedral angles, curvature, and shape diameter and the output is a probability for an edge to be a segmentation boundary. Since the predicted probabilities over the mesh does not correspond to closed smooth boundaries, a thinning and an active contour model \cite{Kass:1988:SAC} are used as post-processing to produce the final segmentations.

\paragraph*{Transductive segmentation.} Another way to formulate the shape segmentation problem is to group patches on a mesh such that the segment similarity is maximized between the resulting segments and the provided segments in the training database. The segment similarity can be measured as the reconstruction cost of the resulting segment from the training ones. The grouping of patches can be solved as an integer programming problem \cite{Xu:2014:TSS}.

\paragraph*{Shape segmentation from labeled images.} Instead of using labeled training shapes for supervised shape segmentation, an alternative source of training data can come in the form of segmented and labeled images, as demonstrated by Wang et al. \cite{Wang:2013:PAS}. Given an input 3D shape, this method first renders 2D binary images of it from different viewpoints. Each binary image is used to retrieve multiple training segmented and labeled images from an input database based on a bi-class Hausdorff distance measure. Each retrieved image is used to perform label transfer to the 2D shape projections. All labeled projections are then back-projected onto the input 3D model to compute a labeling probability map. The energy function for segmentation is formulated by using this probability map in the unary term expressed per face or point, while dihedral angles and Euclidean distances are used in the pairwise term.

\subsection{Semi-supervised shape segmentation}

\paragraph*{Entropy regularization.} The parameters $\theta$ of Equation \ref{eqn:CRFEnergy} can be learned not only from the training labeled shapes, but also from the unlabeled shapes \cite{Lv:2012:SMS}. The idea is that learning should maximize the likelihood function of the parameters over the labeled shapes, and also minimize the entropy (uncertainty) of the classifier over the unlabeled shapes (or correspondingly maximize the negative entropy). The idea is that minimizing the entropy over unlabeled shapes encourages the algorithm to find putative labelings for the unlabeled data \cite{Jiao:2006:SCR}. However, it is generally hard to strike a balance between the likelihood and entropy terms.

\paragraph*{Metric embedding and active learning.} A more general formulation for semi-supervised segmentation was presented in \cite{Wang:2012:ACS}.
Starting from a set of shapes that are co-segmented in an unsupervised manner \cite{Sidi:2011:CS}, the user interactively adds two types of constraints: ``must-link'' constraints, which specify that two patches (super-faces) should belong to the same cluster, and ``cannot-link'' constraints which specify that two patches  must be in different clusters. These constraints are used to perform constrained clustering in an embedded feature space of super-faces coming from all the shapes of the input dataset. The key idea is to transform the original feature space, such that super-faces with ``must-link'' constraints come closer together to form a cluster in the embedded feature space, while super-faces with ``cannot-link'' constraints move away from each other. To minimize the effort required from the user, the method suggests the user pairs of points in feature space that when constrained are likely to improve the co-segmentation.  The suggestions involve points that are far from their cluster centers, and have a low confidence of belonging to their clusters.

\paragraph*{Template fitting.} A different form of partial supervision can come in the form of part-based templates. Kim et al.'s method \cite{Kim:2013:lpt} allows users to specify or refine a few templates made out of boxes representing expected parts in an input database. The boxes iteratively fit to the shapes of a collection through simultaneous alignment, surface segmentation and point-to-point correspondences estimated between each template and each input shape. Alternatively, the templates can be inferred automatically from the shapes of the input collection without human supervision based on single shape segmentation heuristics. Optionally, the user can refine and improve these estimated templates. From this aspect, Kim et al.'s method can run in either semi-supervised or unsupervised method. It was also the first method to handle segmentation and correspondences in collections with size in the order of thousands of shapes.

\subsection{Unsupervised segmentation}

Unsupervised data-driven shape segmentation techniques fall into two categories: clustering based techniques and matching based techniques. In the following, we highlight the key idea of each type of approaches.

\paragraph*{Clustering} based techniques are adapted from supervised techniques. They compute feature descriptors on points or faces. Clustering is performed over all points/faces over all shapes. Each resulting cluster indicates a consistent segment across the input shapes. The promise of the clustering based approach is that when the number of shapes becomes large, the sampling density in the clustering space becomes dense enough, so that certain statistical assumptions are satisfied, e.g., diffusion distances between points from different clusters is significantly larger than those between points within each cluster. When these assumptions are satisfied, clustering based approach can produce results that are comparable to supervised techniques (c.f.~\cite{Hu:2012:CSS}). In addition, clustering method being employed play an important role in the segmentation results. In~\cite{Sidi:2011:CS}, the authors utilize spectral clustering to perform clustering. In~\cite{Hu:2012:CSS}, the authors employ subspace clustering, a more advanced clustering method, to obtain improved results.

Another line of unsupervised methods pursues clustering of parts. In~\cite{Xu:2010:SCS}, the authors perform co-analysis over a set of shapes via factoring out the part scale variation by grouping the shapes into different styles, where style is defined by the anisotropic part scales of the shapes. In~\cite{van-Kaick:2013:CHA}, the authors introduce unsupervised co-hierarchical analysis of a set of shapes. They propose a novel cluster-and-select scheme for selecting representative part hierarchies for all shapes and grouping the shapes according to the hierarchies. The method can be used to compute consistent hierarchical segmentation for the input set.

\begin{figure}[t!]
\centering
    \includegraphics[width=1.0\columnwidth]{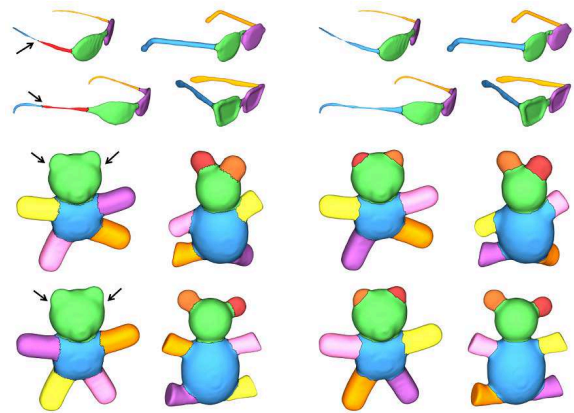}
    \vspace{-0.5cm}
    \caption{Comparison of single-shape segmentation (left) and joint shape segmentation (right) on models from the PSB benchmark~\cite{Chen:2009:BMS}. Each segmentation on the left was produced by the top-performing algorithm in the benchmark for that shape. The segmentations on the right were produced
by~\cite{Huang:2011:JSS}, which jointly optimized segmentations and correspondences across the entire dataset.}
    \vspace{-0.5cm}
    \label{fig:huang_siga11_jss}
\end{figure}

\paragraph*{Matching} based methods~\cite{Golovinskiy:2009:CS,Huang:2011:JSS,Wang:2013:FMap,Huang:2014:FMN} build maps across shapes and utilize these maps to achieve consistency of segmentations. As shown in Figure~\ref{fig:huang_siga11_jss}, this strategy allows us to identify meaningful parts despite the lack of strong geometric cues on a particular shape. Likewise, the approach is able to identify coherent single parts even when the geometry of the individual shape suggests the presence of multiple segments. A challenge here is to find a suitable shape representation so that maps across diverse shapes are well-defined. In~\cite{Huang:2011:JSS}, Huang et al. introduce an optimization strategy that jointly optimizes shape segmentations and maps between optimized segmentations. Since the maps are defined at the part-level, this technique is suitable for heterogeneous shape collections. Experimentally, it generates comparable results with supervised method~\cite{Kalogerakis:2010:LMS} on the Princeton segmentation benchmark.  Recently, Huang et al.\cite{Huang:2014:FMN} formulates the same idea under the framework of functional maps~\cite{Ovsjanikov:2012:FMF} and gain improved segmentation quality and computational efficiency.

\section{Joint Shape Matching}
\label{sec:matching}

Another fundamental problem in shape analysis is shape matching, which finds relations or maps between shapes. These maps allow us to transfer information across shapes and aggregate information from a collection of shapes for a better understanding of individual shapes (e.g., detecting shared structures such as skeletons or shape parts). They also provide a powerful platform for comparing shapes (i.e., with respect to different measures and at difference places). As we can see from other sections, shape maps are widely applied in shape classification and shape exploration as well.

So far most existing research in shape matching has focused on matching pairs of shapes in isolation. We refer to~\cite{van-Kaick:2010:SSC} for a survey and to~\cite{Leordeanu:2005:SM,Lipman:2009:MVC,van-Kaick:2010:SSC,Ovsjanikov:2010:OPM,Kim:2011:BIM,Ovsjanikov:2012:FMF} for recent advances. Although significant progress has been made,
state-of-the-art techniques are limited to shapes that similar to each other. On the other hand, they tend to be insufficient for shapes that undergo large geometric and topological variations.

The availability of large shape collections offers opportunities to address this issue. Intuitively, when matching two dissimilar shapes, we may utilize intermediate shapes to transfer maps. In other words, we can build maps between similar shapes, and use the composite maps to obtain maps between less similar shapes. As we will see shortly, this intuition can be generalized to enforcing a cycle-consistency constraint, namely composite maps along cycles should be identity map or the composite map between two shapes is path-independent.
In this section, we discuss joint shape matching techniques that take a shape collection and initial noisy maps computed between pairs of shapes as input, and output improved maps across the shape collection.

\subsection{Model Graph and Cycle-Consistency}

To formulate the joint matching problem, we consider a model graph $\set{G} = (\set{S}, \mathcal{E})$ (c.f.~\cite{Huber:2002:Thesis}). The vertex set $\set{S} = \{S_1, \cdots, S_{n})\}$ consists of the input shapes. The edge set $\set{E}$ characterizes the pairs of shapes that are selected for performing pair-wise matching. For small-scale datasets, we typically match all pairs of shapes. For large-scale datasets, the edge set usually connects shapes that are similar according to a pre-defined shape descriptor~\cite{Kim:2012:FC,Huang:2013:FSL}, thus generating a sparse shape graph.

The key component of a joint matching algorithm is to utilize the so-called cycle-consistency constraint. Specifically speaking, if all the maps in $\mathcal{G}$ are correct, then composite maps along any loops should be identity maps. This is true for maps that are represented as transformations (e.g., rotations and rigid/affine transformations), or full point-wise maps that can be described as permutation matrices). We can easily modify the constraint to handle partial maps, namely each point, when transformed along a loop, either disappears or goes back to the original point (See \cite{Huang:2014:FMN} for details).

The cycle-consistency constraint is useful because the initial maps, which are computed between pairs of shapes in isolation, are not expected to satisfy the cycle consistency constraint. On the other hand, although we do not know which maps or correspondences are incorrect, we can detect inconsistent cycles. These inconsistent cycles provide useful information for us to detect incorrect correspondences or maps, i.e., an inconsistent cycle indicates that at least one of the participating maps or correspondences is incorrect. To turn this observation into algorithms, one has to formulate the cycle-consistency constraint properly. Existing works in data-driven shape matching fall into two categories: combinatorial techniques and matrix recovery based techniques.  The reminder of this section provides the details.

\begin{figure}[t]
\centering
    \includegraphics[width=1.0\columnwidth]{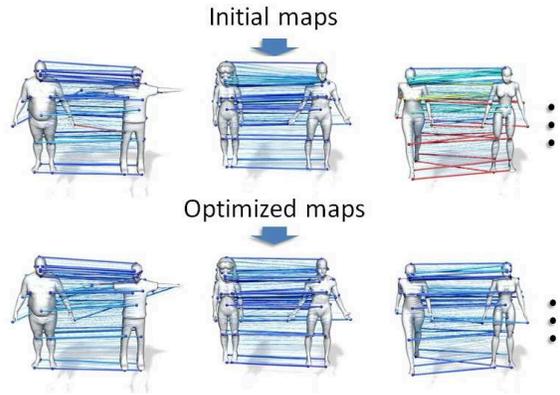}
    \vspace{-0.4cm}
    \caption{Joint shape matching takes as input maps computed between pairs of shapes in isolation and utilizes the cycle-consistency constraint to improve shape maps. This figure shows the result of Huang et al.~\cite{Huang:2014:FMN}, which performs joint shape matching under the functional map setting.}
    \vspace{-0.5cm}
    \label{fig:huang_sig14_jsm}
\end{figure}

\subsection{Combinatorial Techniques}

\para{Spanning tree optimization.} Earlier works in joint matching aim at finding a spanning tree in the model graph. In~\cite{Goldberg:2004:GAA,Huang:2006:RFO}, the authors propose to use the maximum spanning tree (MST) of the model graph. However, this strategy can easily fail since a single incorrect edge in the MST may break the entire matching result. In the seminal work~\cite{Huber:2002:Thesis}, Huber showed that finding the best spanning tree maximizing the number of consistent edges is NP-hard. Although finding the best spanning tree is not tractable, Huber introduced several local operations for improving the score of spanning trees. However, these approaches are generally limited to small-scale problems so that the search space can be sufficiently explored.

\para{Inconsistent cycle detection.} Another line of approaches~\cite{Zach:2010:DVR,Roberts:2011:SFM,Nguyen:2011:CSM} applies global optimization to select cycle-consistent maps. These approaches are typically formulated as solving constrained optimization problems, where objective functions encode the scores of selected maps, and constraints enforce the consistency of selected maps along cycles. The major advantage of these approaches is that the correct maps are determined globally. However, as the cycle consistency constraint needs to apportion blame along many edges on a cycle, the success of these approaches relies on the assumption that correct maps are dominant in the model graph so that the small number of bad maps can be identified through their participation in many bad cycles.

\para{MRF formulation.} Joint matching may also be formulated as solving a second order Markov Random Field (or MRF)~
\cite{Cho:2010:JPS,Cho:2010:TPT,Crandall:2011:DOL,Huang:2012:OAE}. The basic idea is to sample the transformation/deformation space of each shape to obtain a candidate set of transformation/deformation samples per shape. Joint matching is then formulated as optimizing the best sample for each shape. The objective function considers initial maps. Specifically, each pair of samples from two different shapes would generate a candidate map between them. The objective function then formulates second-order potentials, where each term characterize the alignment score between these candidate maps and the initial maps~\cite{Huang:2013:FSL,Huang:2012:OAE}.

The key challenge in the MRF formulation is generating the candidate samples for each shape. The most popular strategy is to perform uniform sampling~\cite{Crandall:2011:DOL,Huang:2013:FSL}, which works well when the transformation space is low-dimensional. To apply the MRF formulation on high-dimensional problems, Huang et al.~\cite{Huang:2012:OAE} introduce a diffusion-and-sharpening strategy. The idea is to diffuse the maps among the model graph to obtain rich samples of candidate transformations or correspondences and then perform clustering to reduce the number of candidate samples.

\subsection{Matrix Based Techniques}

A recent trend in map computation is to formulate joint map computation as inferring matrices~\cite{Singer:2011:VDM,Kim:2012:FC,Huang:2012:OAE,journals/corr/abs-1211-2441,Huang:2013:SDP,Chen:2014:SDP,Huang:2014:FMN}. The basic idea is to consider a big map collection matrix
$$
\vec{X} = \left [
\begin{array}{cccc}
\vec{X}_{11} & \vec{X}_{12} & \cdots & \vec{X}_{1n} \\
\vec{X}_{21} & \vec{X}_{22} & \cdots & \vec{X}_{2n} \\
\vdots & \ddots & \cdots & \vdots \\
\vec{X}_{21} & \cdots & \cdots & \vec{X}_{nn}
\end{array}
\right],
$$
where each block $\vec{X}_{ij}$ encodes the map from shape $S_i$ to shape $S_j$. In this matrix representation, the cycle-consistency constraint can be equivalently described as simple properties of $\vec{X}$, i.e., depending on the types of maps, $\vec{X}$ is either positive semidefinite or low-rank (c.f.~\cite{Huang:2013:SDP,Huang:2014:FMN}). In addition, we may view the initial pair-wise maps as noisy measurements of the entries of $\vec{X}$. Based on this perspective, we can formulate joint matching as matrix recovery from noisy measurements of its entries.

\para{Spectral techniques.} The initial attempts in matrix recovery are spectral techniques and their variants~\cite{Singer:2011:VDM,Kim:2012:FC,Wang:2013:FMap}. The basic idea is to consider the map collection $\vec{X}^{\textup{input}}$ that encodes initial maps in its blocks. Then the recovered matrix is given by $\vec{X} = \vec{U}\Sigma \vec{V}^{T}$, where $\vec{U}, \Sigma, \vec{V}$ are given singular value decomposition (or SVD) of $\vec{X}^{\textup{input}}$. Various methods have added heuristics on top of this basic procedure. For example, Kim et al.~\cite{Kim:2012:FC} use the optimized maps to recompute initial maps.

This SVD strategy can be viewed as matrix recovery because $\vec{X}$ is equivalent to the optimal low-rank approximation of $\vec{X}^{\textup{input}}$ (with given rank) under the matrix Frobenius norm. However, as the input maps may contain outliers, employing the Frobenius norm for matrix recovery is sub-optimal. Moreover, it is hard to analyze these techniques, even in the very basic setting where maps are given by permutation matrices~\cite{conf/nips/PachauriKS13}.

\para{Point-based maps.} In a series of works, Huang and coworkers~\cite{Huang:2013:SDP,Chen:2014:SDP,cg-ssrmrf-14} consider the case of point-based maps and develop joint matching algorithms that admit theoretical guarantees. The work of~\cite{Huang:2013:SDP} considers the basic setting of permutation matrix maps and proves the equivalence between cycle-consistent maps and the low-rank or positive semi-definiteness of the map collection matrix. This leads to a semidefinite programming formulation for joint matching. In particular, L1 norm is used to measuring the distance between the recovered maps and the initial maps. The authors provide exact recovery conditions, which state that the ground-truth maps can be recovered if the percentage of incorrect correspondences in the input maps is below a constant. In a followup work, Chen et al.~\cite{Chen:2014:SDP} extends it to partial maps and provide a better analysis in the case where incorrect correspondences in the input maps are random. The computational issue is addressed in~\cite{cg-ssrmrf-14}, which employs alternating direction of multiplier methods for optimization.

\para{Rotations and functional maps.} Maps that are represented by general matrices (e.g., rotations or functional maps) can also be handled in a similar fashion. In~\cite{journals/corr/abs-1211-2441}, Wang and Singer consider the case of rotations between objects. Their formulation is similar to~\cite{Huang:2013:SDP} but utilize a L1 Frobenius norm for measuring the distance between initial rotations and recovered rotations. Recently, Huang et al.~\cite{Huang:2014:FMN} extend the idea to functional maps. The major difference between functional maps and point-based maps or rotations is that the map collection matrix is no-longer symmetric. Thus, their method is formulated to recover low-rank matrices.

\subsection{Discussion and Future Directions}

The key to a joint shape matching algorithm is to have a proper formulation of the cycle-consistency constraint. We have witnessed the evolution from earlier works on combinatorial search and detecting inconsistent cycles to more recent works on spectral techniques, MRF based methods and matrix recovery techniques. In particular, matrix recovery techniques admit theoretical guarantees. They provide fundamental understanding of why joint shape matching can improve from isolated pair-wise matching.

One future direction is to integrate pair-wise matching and joint matching into one optimization problem. Since the major role of joint matching is to remove the noise presented in pair-wise matching, it makes sense to perform them together. Such unified approaches have the potential to further improve from decomposed approaches (i.e., from pair-wise to joint). The technical challenge is to find map representations so that pair-wise matching and map consistency can be formulated in the same framework.


\section{Data-Driven Shape Reconstruction}
\label{sec:recon}

Reconstructing geometric shapes from physical objects is a fundamental problem in geometry processing. The input to this problem is usually a point cloud produced by aligned range scans, which provides an observation of an object. The goal of a shape reconstruction algorithm is to convert this point cloud into a high-quality geometric model. In practice, the input point cloud data is noisy and incomplete, thus the key to a successful shape reconstruction algorithm is formulating appropriate shape priors. Traditional shape reconstruction algorithms usually utilize generic priors, such as surface smoothness \cite{Diebel:2005:BMP}, and typically assume that the input data captures most of the object's surface.  To handle higher degree of noise and partiality of the input data, it is important to build structural shape priors.

Data-driven techniques tackle this challenge by leveraging shape collections to learn strong structural priors from similar objects, and use them to reconstruct high-quality 3D models. Existing approaches fall into two categories, based on how they represent the shape priors: \textsl{parametric} and \textsl{non-parametric}. The former usually builds a low-dimensional parametric representation of the underlying shape space, learning the representation from exemplars and enforcing the parameterization when reconstructing new models. Parametric methods typically require building correspondences across the exemplar shapes.
In contrast, non-parametric methods directly operate on the input shapes by copying and deforming existing shapes or shape parts, which are designed for shapes with large variations, such as man-made objects.

\subsection{Parametric Methods}

\begin{figure}[t]
\centering
    \includegraphics[width=1.0\columnwidth]{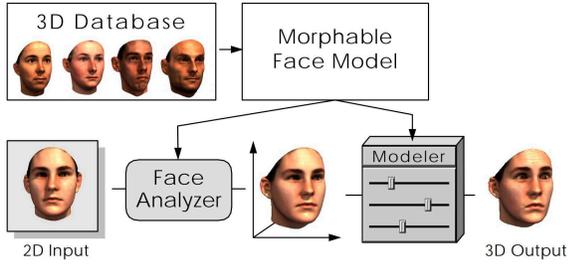}
    \vspace{-0.4cm}
    \caption{Derived from a dataset of prototypical 3D scans of faces, the morphable face model contributes to two main steps in face manipulation: (1) deriving a 3D face model from a novel image, and (2) modifying shape and texture in a natural way~\cite{Blanz:1999:MMS}.}
    \vspace{-0.5cm}
    \label{fig:blanz_sig99_face}
\end{figure}

\para{Morphable face.} A representative work in parametric data-driven shape reconstruction is the morphable face model~\cite{Blanz:1999:MMS},
which is designed for reconstructing 3D textured faces from photos and scans. The model is learned from a dataset of prototypical 3D shapes of faces, and the model can then be used to derive a 3D face model from a novel image and to modify shape and texture in a natural way (See Figure\ref{fig:blanz_sig99_face}).

In particular, the morphable face model represents the geometry of a face with shape-vector $S = (\mathbf{p}_1^{T}, \cdots, \mathbf{p}_{n}^{T})^{T}) \in \mathbb{R}^{3n})$, that contains the 3D coordinates of its $n$ vertices. Similarly, it encodes the texture of a face by a texture-vector $T = (\vec{c}_1^{T}, \vec{c}_2^{T}, \cdots, \vec{c}_{n}^{T}) \in \mathbb{R}^{3n}$, that contains the RGB color values of the corresponding vertices. A morphable face model is then constructed using a database of $m$ exemplar faces, each represented by its shape-vector $S_i$ and $T_i$. In~\cite{Blanz:1999:MMS} the exemplar faces are constructed by matching a template to scanned human faces.

The morphable face model uses Principal Component Analysis (PCA) to characterize the shape space. A new shape and its associated texture are given by
$$
\vec{S}_{\textup{mod}} = \overline{\vec{S}} + \sum\limits_{i=1}^{m-1}\alpha_i \vec{s}_i, \quad \vec{T}_{\textup{mod}} = \overline{\vec{T}} + \sum\limits_{i=1}^{m-1}\beta_i \vec{t}_i,
$$
where $\overline{\vec{S}}$ and $\overline{\vec{T}}$ are the mean-shape and mean-texture, respectively, and $\vec{s}_i$ and $\vec{t}_i$ are eigenvectors of covariance matrices. $\alpha_i$ and $\beta_i$ are coefficients. PCA also gives probability distributions over coefficients. The probability for coefficients $\alpha_i$ is given by
$$
p(\{\alpha_i\}) \sim \exp\left(-\frac{1}{2}\sum\limits_{i=1}^{m-1}(\alpha_i/\sigma_i)^2\right),
$$
with $\sigma_i^2$ being the eigenvalues of the shape covariant matrix $C_{S}$ (the probability $p(\{\beta_i\})$ is computed in a similar way).

With this morphable face model, reconstruction of textured models can be posed as a small-scale non-linear optimization problem. For example, given a 2D image of a human face $I_{\textup{input}}$, one can reconstruct the underlying textured 3D model by searching for a similar rendered face $I(\{\alpha_i\},\{\beta_i\}, p)$, parameterized by the shape and texture coefficients $\alpha_i$ and $\beta_i$, and the rendering parameters $p$ (e.g., camera configuration, lighting parameters). The optimization problem is formulated as minimizing a data term, which measures the distance between the input image and the rendered image, and regularization terms that are learned from exemplar faces. The success of the morphable model relies on low-dimensionality of the solution space, thus this method was applied to several other data sets where this assumption holds, such as human bodies and poses.


\begin{figure}[t]
\centering
    \includegraphics[width=1.0\columnwidth]{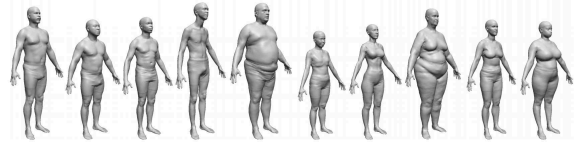}
    \vspace{-0.4cm}
    \caption{Parameterizing the variation in human shapes can be used to synthesize new individuals or edit existing ones~\cite{Allen:2003:SHB}.}
    \vspace{-0.6cm}
    \label{fig:allen_sig03_human}
\end{figure}

\para{Morphable human bodies.} Allen et al.~\cite{Allen:2003:SHB} generalize morphable model to characterize human bodies (Figure ~\ref{fig:allen_sig03_human}). Given a set of $250$ scanned human bodies, the method first performs non-rigid registration to fit a hole-free, artist-generated mesh (template) to each of these scans. The result is a set of mutually consistent parameterized shapes based on the corresponding vertex positions originating from the template. Similar to~\cite{Blanz:1999:MMS}, the method employs PCA to characterize the shape space, which enables applications in shape exploration, synthesis and reconstruction.

In addition to variations in body shapes, human models exhibit variations in poses.  The SCAPE model (Shape Completion and Animation for PEople)~\cite{SCAPE:2005} addresses this challenge by learning separate models of body deformation -- one accounting for variations in poses and one accounting differences in body shapes among humans. The pose deformation component is acquired from a set of dense 3D scans of a single person in multiple poses. A key aspect of the pose model is that it decomposes deformation into a rigid and a non-rigid component. The rigid component is modeled using a standard skeleton system. The non-rigid component, which captures remaining deformations such as flexing of the muscles, associates each triangle with a local affine transformation matrix. These transformation matrices are learned from exemplars using a joint regression model. 
In~\cite{Hasler:2009:SSR}, Hasler et al. introduce a unified model for parameterizing both shapes and poses. The basic idea is to consider the relative transformations between all pairs of neighboring triangles. These transformation matrices allow us to reconstruct the original shape by solving a least square problem. In this regard, each shape is encoded as a set of edge-wise transformation matrices, which are fit into the PCA framework to obtain a statistical model of human shapes. The model is further used to estimate shapes of dressed humans from range scans~\cite{Hasler:2009:TSE}.

Recent works on statistical human shape analysis focus on combing learned shape priors with sparse observations and special effects. In \cite{Tsoli:2014:BLS}, the authors introduce an approach that reconstruct high-quality shapes and poses from a sparse set of markers. The success of this approach relies on learning meaningful shape priors from a database consists of thousands of shapes. In~\cite{Loper:SIGASIA:2014}, the authors study how to understand human breathing from acquired data.



\para{Data-driven tracking.} Another problem in shape reconstruction is object tracking, which aims at creating and analyzing dynamic shapes and/or poses of physical objects. Successful tracking techniques (e.g.,~\cite{Weise:2009:FLF,Weise:2011:RPF,Li:2013:RFA,Cao:2013:SRR,Cao:2014:DDE}) typically utilize parametric shape spaces. These reduced shape spaces provide shape priors that improve both the efficiency and robustness of the tracking process. The way to utilize and construct shape spaces vary in different settings, and are typically tailored to the specific problem setting. Weise et al.~\cite{Weise:2009:FLF} utilize a linear PCA subspace trained with a very large set of pre-processed facial expressions. This method requires an extended training session with a careful choice of facial action units. In addition, the learned face model is actor-specific. These restrictions are partially resolved in~\cite{Li:2010:EFR}, which introduces an example-based blendshape optimization technique, involving only a limited number of random facial expressions.
In~\cite{Weise:2011:RPF}, the authors combine both blendshapes and data-driven animation priors to improve the tracking performance. In a recent work, Li et al.~\cite{Li:2013:RFA} employs adaptive PCA to further improve the tracking performance on nuanced emotions and micro-expression. The key idea is to combine a general blendshape PCA model and a corrective PCA model that is updated on-the-fly. This corrective PCA model captures the details of the specific actor and missing deformations from the initial blendshape model.

\vspace{-.2cm}

\subsection{Non-Parametric Methods}

Parametric methods require canonical domains to characterize the shape space, which have been so far demonstrated in domains of organic shapes,  such as body shapes or faces.
In this section, we discuss another category of methods that have shown the potential to handle more diverse shape collections.

Generally speaking, a non-parametric data-driven shape reconstruction method utilizes a collection of relevant shapes and combines three phases, i.e., a query phase, a transformation phase and a assembly phase. Existing methods differ in how the input shape collection is preprocessed and how these phases are performed.

\paragraph*{Example-based scan completion.}
Pauly et al.~\cite{Pauly:2005:ESC} introduce one of the first non-parametric systems. As shown in~\cite{Pauly:2005:ESC}, the method takes an input point cloud and a collection of complete objects as input. The reconstruction procedure reveals all three phases described above. The first phase determines a set of similar objects. The retrieval phase combines both text-based search, PCA signatures and is refined by rigid alignment. The second step performs non-rigid alignment between the retrieved shapes and the input point cloud. This step partitions the input point cloud into a set of patches, where each patch is associated with one retried shape (via the corresponding region). The final phase merges the corresponding regions into a unified shape.

Nan et al.~\cite{Nan:2012:SAC} introduce a similar system for indoor scene reconstruction. Given an input point cloud of an indoor scene that consists of a set of objects with known categories, the method searches in a database of 3D models to find matched objects and then deforms them in a non-rigid manner to fit the input point cloud. Note that this method treats complete 3D objects as building blocks, so the final reconstruction does not necessarily reflect the original scene.

In contrast to considering entire 3D shapes, Gal et al.~\cite{Gal:2007:SRU} utilizes a dictionary of local shape priors (defined as patches) for shape reconstruction. The method is mainly designed for enhancing shape features, where each region of an input point cloud is matched to a shape patch in the database. The matched shape patch is then used to enhance and rectify the local region. Recently, Mattausch et al.~\cite{MATTAUSCH:2014:ODC} introduce a patch-based reconstruction system for indoor scenes. Their method considers recognizing and fitting planar patches from point cloud data.

Shen and coworkers~\cite{Shen:2012:SRP} extends the idea for single object reconstruction, by assembling object parts. Their method utilizes consistently segmented 3D shapes as the database. Given a scan of an object, it recursively search parts in the database to assemble the original object. The retrieval phase considers both the geometric similarity between the input and the retrieved parts and the part compatibility learned from the input shapes.

\paragraph*{Data-driven SLAM.}
Non-parametric methods have also found applications in reconstructing temporal geometric data (e.g., the output of the Kinect scanner). A notable technique is simultaneous localization and mapping (or SLAM) method, which jointly estimates the trajectory of the scanning device and the geometry of the environment. In this case, shape collections serve us priors for the objects in the environment, which could be used to train object detectors. For example, the SLAM++ system proposed by Salas-Moreno et al.~\cite{Salas-Moreno:2013:SLAM} trained domain specific object detectors from shape collections. The learned detectors are integrated inside the SLAM framework to recognize and track those objects. Similarly Kim et al.~\cite{Kim:2012:AIE} use learned object models to reconstruct dense 3D models from a single scan of an indoor scene. More recently, Sun et al.~\cite{Sun:2014:SS} introduce 3D sliding window object detector with improved performance and broader range of objects.

\para{Shape-driven reconstruction from images.} Recently, there is a growing interest in reconstructing 3D objects directly from images (e.g.,~\cite{Xu:2011:PMO,Kholgade:2014:OMS,Aubry14,Su:2014:EID}). This problem introduces fundamental challenges in both querying similar objects and deforming objects/parts to fit the input object. In terms of searching similar objects, successful methods typically render objects in the database from a dense of viewpoints and pick objects, where one view is similar to the input image object. Since the depth information is missing from the image object, it is important to properly regularize 3D object transformations. Since otherwise a 3D object maybe deformed arbitrarily even though its projection on the image domain matches the image object. Most existing techniques consider rigid transformations or user-specified deformations~\cite{Xu:2011:PMO}. In a recent work, Su et al.~\cite{Su:2014:EID} propose to learn meaningful deformations of each shape from its optimal deformations to similar shapes.

\section{Data-driven Shape Modeling and Synthesis}
\label{sec:modeling}

So far, the creation of detailed three-dimensional content remains a tedious task confined with skilled artists.
3D content creation has been a major bottleneck hindering the development of ubiquitous 3D graphics.
Thus, providing easy-to-use tools for casual and novice users to design and create 3D models has been a key
challenge in computer graphics. To address this challenge, current literature has been focused on two main directions, i.e.,
intelligent interfaces for interactive shape modeling and smart models for automated model synthesis.
%
The former strives to endow modeling interfaces with higher-level understanding of the structure and semantics
of 3D shapes, allowing the interface to reason around the incomplete shape being modeled. The latter direction focuses on developing data-driven models to synthesize new shapes automatically.
The core problem is to learn generative shape models from a set of exemplars (e.g., probability distributions, fitness functions, functional constraints etc) so that the synthesized shapes are plausible and novel.
It can be seen that both of the two paradigms depend on data-driven modeling of shape structures and semantics.
With the availability of large 3D shape collections, data-driven approach seems a promising breakthrough to the content creation bottleneck.

\begin{figure}[t]
\centering
    \includegraphics[width=1.05\columnwidth]{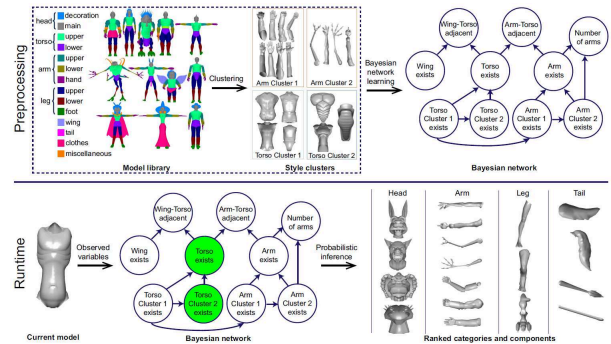}
    \vspace{-0.4cm}
    \caption{
    Given a library of models, a Bayesian network encoding semantic and geometric relationships among shape parts
    is learned~\cite{Chaudhuri:2011:prabm} (top). The modeling process (bottom) performs probabilistic inference in the learned
    Bayesian network to generate ranked lists of category labels and components within each category, customized for the currently assembled model.}
    \vspace{-0.6cm}
    \label{fig:segmentation_labeling}
\end{figure}

\subsection{Interactive Shape Modeling and Editing}
\label{sec:interactive_modeling}
Interactive 3D modeling software (3DS Max, Maya, etc.) provide the artists with a big set of powerful tools
for creating and editing very detailed 3D models,
which are, however, often onerous to harness for non-professional users.
For casual users, more intuitive modeling interfaces with certain intelligence are preferred. Below we discuss such methods for assembly-based modeling and guided shape editing.


\paragraph*{Data-driven part assembly.}
Early works on 3D modeling based on shape sets are primarily driven by the purpose of \emph{content reuse}
in part-assembly based modeling approaches.
The seminal work of modeling by example~\cite{Funkhouser:2004:MBE} presents a pioneering system
of shape modeling by searching a shape database for parts to reuse in the construction of new shapes.
Kraevoy et al.~\shortcite{Kreavoy:2007:MIC} describe a system for shape creation
via interchanging parts between a small set of compatible shapes.
Guo et al.~\cite{Guo:2014:CG} propose assembly-based creature modeling guided by a shape grammar.

Beyond content reuse through database queries or hand-crafted rules, Chaudhuri and Koltun~\shortcite{Chaudhuri:2010:ddsc} propose a data-driven technique
for suggesting the modeler with shape parts that can potentially augment the current shape being built.
Such part suggestions are generated through retrieving a shape database based on partial shape matching.
Although this is a purely geometric method without accounting for the semantics about shape parts,
it represents the first attempt on utilizing shape database to \emph{augment the modeling interface}.
Later, Chaudhuri et al.~\shortcite{Chaudhuri:2011:prabm} show that the incorporation of
semantic relationships increases the relevance of presented parts.
Given a repository of 3D shapes, the method learns a probabilistic graphical model encoding semantic and geometric
relationships among shape parts. During modeling, inference in the learned Bayesian network
is performed to produce a relevance ranking of the parts.

A common limitation of the above techniques is that they do not provide a way to directly express a high-level design goal (e.g. ``create a cute toy''). Chaudhuri et al.~\shortcite{Chaudhuri:2013:ACC} proposed a method that learns semantic attributes for shape parts that reflect the high-level intent people may have for creating content in a domain (e.g. adjectives such as ``dangerous'', ``scary'' or ``strong'') and ranks them according to the strength of each learned attribute (Figure \ref{fig:attribit}). During an interactive session, the user explores and modifies the strengths of semantic attributes to generate new part assemblies.

3D shape collections can supply other useful information, such as contextual and spatial relationships between shape parts,
to enhance a variety of modeling interfaces.
Xie et al.~\cite{Xie:2013:S2D} propose a data-driven sketch-based 3D modeling system.
In the off-line learning stage, a shape database is pre-analyzed to extract the contextual information among parts.
During the online stage, the user designs a 3D model by progressively sketching its parts and retrieving and assembling
shape parts from the database. Both the retrieval and assembly are assisted by the
precomputed contextual information so that more relevant parts can be returned and selected parts can be automatically placed.
Inspired by the ShadowDraw system~\cite{Lee:2011:SD}, Fan et al.~\cite{Fan:2013:MBD} propose 3D modeling by drawing with
data-driven shadow guidance. The user's strokes are used to query a 3D shape database for generating the shadow image,
which in turn can guide the user's drawing. Along the drawing, 3D candidate parts are retrieved for assembly-based modeling.

\paragraph*{Data-driven editing and variation.}
The general idea of data-driven shape editing is to learn from a collection of closely related shapes a model
that characterize the plausible variation or deformation of the shapes, and use the learned model to constrain
the user's edit to maintain plausibility.
For organic shapes, such as human faces~\cite{Blanz:1999:MMS,Chen:2014:FED} or bodies~\cite{Allen:2003:SHB},
parametric models can be learned from a shape set characterizing its shape space.
Such parametric models can be used to edit the shapes through exploring the shape space with the set of parameters.

An alternative approach is the analyze-and-edit paradigm that is widely adopted to first extract the structure from the input shape and then try to preserve the structure through constraining the editing~\cite{Gal:2009:IAA}.
Instead of learning structure from a single shape, which usually relies on prior-knowledge, Fish et al.~\cite{Fish:2014:MR}
learn it from a set of shapes belong to the same family, resulting in a set of geometric distributions characterizing
the part arrangements. These distributions can be used to guide structure-preserving editing, where models can be
edited while maintaining their familial traits.
Yumer et al.~\cite{Yumer:2014:CCH} extract co-constrained handles from a set of shapes for shape deformation.
The handles are generated based on co-abstraction~\cite{Yumer:2012:CSC} of the set of shapes and the deformation co-constraints are learned
statistically from the set.

Based on learned structure from a database of 3D models, Xu et al.~\cite{Xu:2011:PMO} propose photo-inspired 3D object modeling.
Guided by the object in a photograph, the method creates a 3D model as a geometric variation of a candidate model retrieved from the database.
Due to the pre-analyzed structural information, the method addresses the ill-posed problem of 3D modeling from a single 2D image
via structure-preserving 3D warping. The final result is structurally plausible and is readily usable for subsequent editing.
Moreover, the resulting 3D model, although built from a single view, is structurally coherent from all views.

\subsection{Automated Synthesis of Shapes}
\label{sec:synthesis}

\begin{figure}[t]
\centering
    \includegraphics[width=.95\columnwidth]{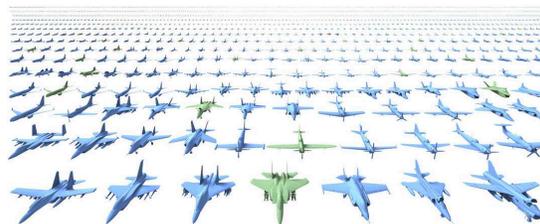}
    \vspace{-0.3cm}
    \caption{
Given a hundred training airplanes (in green), the probabilistic model from \cite{Kalogerakis:2012:PMC} synthesizes several hundreds of new airplanes (in blue).}
    \vspace{-0.6cm}
    \label{fig:kalogerakis_sig12_synthesis}
\end{figure}

Many applications such as 3D games and films require large collections of 3D shapes for populating their environments. Modeling each shape individually can be tedious even with the best interactive tools. The goal of data-driven shape synthesis algorithms is to generate several shapes automatically with no or very little user supervision: user may only provide some preferences or high-level specifications to control the shape synthesis. Existing methods achieve this task by using probabilistic generative models of shapes, evolutionary methods, or learned probabilistic grammars.

\paragraph*{Statistical models of shapes.} The basic idea of these methods is to define a parametric shape space and then fit a probability distribution to the data points that represent the input exemplar shapes. Since the input shapes are assumed to be plausible and desired representatives of the shape space, high-probability areas of the shape space with tend to become associated with new, plausible shape variants. This idea was first explored in the context of parametric models \cite{Blanz:1999:MMS,Allen:2003:SHB}, discussed in Section \ref{sec:recon}. By associating each principal component of the shape space defined by these methods with a Gaussian distribution, this distribution can be sampled to generate new human faces or bodies (Figure \ref{fig:allen_sig03_human}). Since the probability distribution of plausible shapes tend to be highly non-uniform in several shape classes, Talton et al. \cite{Talton:2009:EMC} use kernel density estimation with Gaussian kernels to represent plausible shape variability. The method is demonstrated to generate new shapes based on tree and human body parametric spaces.

Shapes have structure i.e., shapes vary in terms of their type and style, different shape styles have different number and type of parts, parts have various sub-parts that can be made of patches, and so on. Thus, to generate shapes in complex domains, it is important to define shape spaces over structural and geometric parameters, and capture hierarchical relationships between these parameters at different levels. Kalogerakis et al. \cite{Kalogerakis:2012:PMC} (Figure \ref{fig:kalogerakis_sig12_synthesis}) proposed a probabilistic model that represents variation and relationships of geometric descriptors and adjacency features for different part styles, as well as variation and relationships of part styles and repetitions for different shape styles. The method learns the model from a set of consistently segmented shapes. Part and shape styles are discovered based on latent variables that capture the underlying modes of shape variability. Instead of sampling, the method uses a search procedure to assemble new shapes from parts of the input shapes according to the learned probability distribution. Users can also set preferences for generating shapes from a particular shape style, with given part styles or specific parts.

\paragraph*{Set evolution.} Xu et al. \cite{Xu:FDS:2012} developed a method for generating shapes inspired by the theory of evolution in biology. The basic idea of set evolution is to define cross-over and mutation operators on shapes to perform part warping and part replacement. Starting from an initial generation of shapes with part correspondences and built-in structural information such as inter-part symmetries, these operators are applied to create a new generation of shapes. A selected subset from the generation is presented via a gallery to the user who provides feedback to the system by rating them. The ratings are used to define the fitness function for the evolution. Through the evolution, the set is personalized and populated with shapes that better fit to the user. At the same time, the system explicitly maintains the diversity of the population so as to prevent it from converging into an ``elite'' set.

\paragraph*{Learned Shape Grammars.} Talton et al. \cite{Talton:2012:LDP} leverage techniques
from natural language processing to learn probabilistic generative grammars of shapes. The method takes as input a set of exemplar shapes represented with a scene graph specifying parent/child relationships and
relative transformations between labeled shape components. They use Bayesian inference to learn a probabilistic formal grammar that can be used to synthesize novel shapes.

\section{Data-driven Scene Analysis and Synthesis}
\label{sec:scene}
Analyzing and modeling indoor and outdoor environments has important applications in various domains. For example, in robotics it is essential for an autonomous agent to understand semantics of 3D environments to be able to interact with them. In urban planning and architecture, professionals build digital models of cities and buildings to validate and improve their designs. In computer graphics, artists create novel 3D scenes for movies and video games.

\begin{figure}[t]
\centering
    \includegraphics[width=0.99\linewidth]{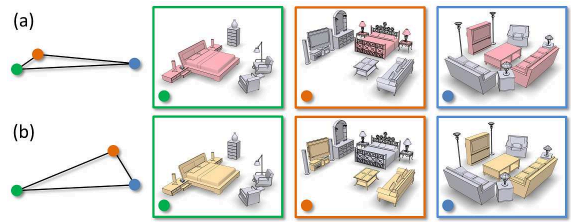}
    \vspace{-0.2cm}
    \caption{
    Scene comparisons may yield different similarity distances (left) depending on the focal points~\cite{Xu:2014:OHSC}.}
    \vspace{-0.4cm}
    \label{fig:xu_sig14_focal}
\end{figure}

Growing numbers of 3D scenes in digital repositories provide new opportunities for data-driven scene analysis, editing, and synthesis. Emerging collections of 3D scenes pose novel research challenges that cannot be easily addressed with existing tools.  In particular, representations created for analyzing collections of single models mostly focus on arrangement and relations between shape parts~\cite{Mitra:2014:SASP}, which usually exhibit less variations than objects in scenes. Capturing scene structure poses a greater challenge due to looser spatial relations and a more diverse mixture of functional substructures.

%

Inferring scene semantics is a long-standing problem in image understanding, with many methods developed for object recognition~\cite{quattoni2009}, classification~\cite{swadzba2010}, inferring spatial layout~\cite{Choi:2013:UIS}, and other 3D information~\cite{Fouhey:2013:DDP} \emph{from a single image}. Previous work demonstrates that one can leverage collections of 3D models to facilitate scene understanding in images~\cite{Satkin:2012:DDS}. In addition, the RGBD scans that include depth information can be used as training data for establishing the link between 2D and 3D for model-driven scene understanding~\cite{Silberman:2012:ISS}. Unfortunately, semantic annotations of images are not immediately useful for modeling and synthesizing 3D scenes, where priors have to be learned from 3D data.

In this section, we cover data-driven techniques that leverage collections of 3D scenes for modeling, editing, and synthesizing novel scenes.


\paragraph*{Context-based retrieval.}
To address large variance in arrangements and geometries of objects in scenes, Fisher et al.~\cite{Fisher:2010:CSM,Fisher:2011:CSR} suggest to take advantage of local context.  One of the key insights of their work is that collections of 3D scenes provide rich information about context in which objects appear. They show that capturing these contextual priors can help in scene retrieval and editing.

Their system takes an annotated collection of 3D scenes as input, where each object in a scene is classified. They represent each scene as a graph, where nodes represent objects and edges represent relations between objects, such as support and surface contact. In order to compare scenes, they define kernel functions for pairs of nodes measuring similarity in object's geometry, and for pairs of edges, measuring similarity in relations of two pairs of objects.  They further define a graph kernel to compare pairs of scenes.  In particular, they compare all walks of fixed length originating at all pairs of objects in both scene graphs, which loosely captures similarities of all contexts in which objects appear~\cite{Fisher:2011:CSR}.  They show that this similarity metric can be used to retrieve scenes. By comparing only paths originated at a particular object, they can retrieve objects for interactive scene editing.


%

\paragraph*{Focal Points.}
Measuring similarity of complex hybrid scenes such as studios composed of bedroom, living room, and dining room poses a challenge to graph kernel techniques since they only measure global scene similarity. Thus, Xu et al.~\shortcite{Xu:2014:OHSC} advocate analyzing salient sub-scenes, which they call focal points, to compare hybrid scenes, i.e., scenes containing multiple salient sub-scenes. Figure~\ref{fig:xu_sig14_focal} shows an example of comparing complex scenes, where
the middle scene is a hybrid one encompassing two semantically salient sub-scenes, i.e., bed-nightstands and TV-table-sofa.
The middle scene is closer to the left one when the bed and nightstands are focused on, and otherwise when the TV-table-sofa combo is the focal point.
Therefore, scene comparison may yield different similarity distances depending on the focal points.

Formally, a focal point is defined as a representative substructure of a scene which can characterize a semantic scene category. That means the substructure should re-occur frequently only within that category. Therefore, focal point detection is naturally coupled with the identification of scene categories via scene clustering. This poses coupled problems of detecting focal points based on scene groups and grouping scenes based on focal points.
These two problems are solved via interleaved optimization which alternates between focal point detection and focal-based scene clustering. The former is achieved by mining frequent substructures and the latter uses subspace clustering, where scene distances are defined in a focal-centric manner.  Inspired by work of Fisher et al.~\cite{Fisher:2011:CSR} scene distances is computed using focal-centric graph kernels which are estimated from walks originating from representative focal points.

\begin{figure}[t] \centering
    \includegraphics[width=0.98\linewidth]{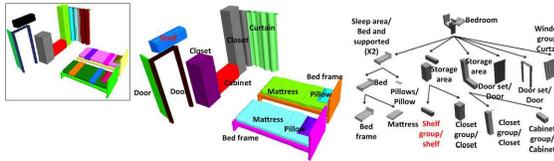}
        \vspace{-0.3cm}
    \caption{
    The algorithm processes raw scene graphs with possible over-segmentation (a) into consistent hierarchies capturing semantic and functional groups (b,c)~\cite{Liu:2014:CCS}.
    }
    \vspace{-0.3cm}
    \label{fig:liu_siga14_sh}
\end{figure}

The detected focal points can be used to organize the scene collection and to support efficient exploration of the collection (see Section~\ref{sec:exploration}). Focal-based scene similarity can be used for novel applications such as multi-query scene retrieval
where one may issue queries consisting of multiple semantically related scenes and wish to retrieve more scenes ``of the same kind''.

\paragraph*{Synthesis.}
Given an annotated scene collection, one can also synthesize new scenes that have similar distribution of objects. The scene synthesis technique of Fisher et al.~\shortcite{Fisher:2012:CSR} learns two probabilistic models from the training dataset: (1) object occurrence, indicating which objects should be placed in the scene, and (2) layout optimization, indicating where to place the objects. Next, it takes an example scene, and then synthesizes similar scenes using the learned priors. It replaces or adds new objects using context-based retrieval techniques, and then optimizes for object placement based on learned object-to-object spatial relations.  Synthesizing example scenes might be a challenging task, thus Xu et al.~\shortcite{Xu:2013:S2S} propose modeling 3D indoor scenes from 2D sketches, by leveraging a database of 3D scenes. Their system jointly optimizes for sketch-guided co-retrieval and co-placement of all objects.

\begin{figure}[t] \centering
    \includegraphics[width=0.99\linewidth]{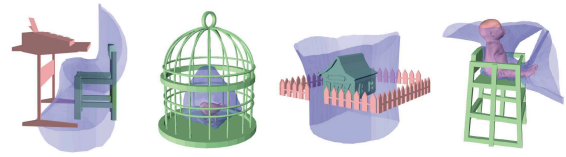}
    \vspace{-0.2cm}
    \caption{
    The interaction bisector surface (in blue) of several two-object scenes~\cite{Zhao:2014:ISU}.
    }
    \vspace{-0.4cm}
    \label{fig:zhao_tog14_ibs}
\end{figure}

\paragraph*{Hierarchical scene annotation.}
All aforementioned applications take an annotated collection of 3D scenes as an input. Unfortunately, most scenes in public repositories are not annotated and thus require additional manual labeling~\cite{Fisher:2012:CSR}. Liu et al.~\shortcite{Liu:2014:CCS} address the challenge of annotating novel scenes. The key observation of their work is that understanding hierarchical structure of a scene enables efficient encoding of functional scene substructures, which significantly simplifies detecting objects and representing their relationships. Thus, they propose a supervised learning approach to estimate hierarchical structure for novel scenes. Given a collection of scene graphs with consistent hierarchies and labels, they train a probabilistic hierarchical grammar encoding the distributions of shapes, cardinalities, and spatial relationships between objects. Such grammar can then be used to parse new scenes: find segmentations, object labels, and hierarchical organization of objects consistent with the annotated collection (see Figure~\ref{fig:liu_siga14_sh}).

\paragraph*{Challenges and opportunities.}
The topic of 3D scene analysis is quite new and there are many open problems and research opportunities.
\emph{The first} problem is to efficiently characterize spatial relationships between objects and object groups. Most existing methods work with bounding box representation which is efficient to process, but not sufficiently informative to characterize object-to-object relationships.
For example, one cannot reliably determine the object enclosure relationship based on a bounding box.
Recently, He et al.~\shortcite{Zhao:2014:ISU} propose to use biologically-inspired bisector surface to characterize the
geometric interaction between adjacent objects and index 3D scenes (Figure~\ref{fig:zhao_tog14_ibs}).
\emph{Second}, most existing techniques heavily rely on expert user supervision for scene understanding. Unfortunately, online repositories rarely have models with reliable object tags. Therefore there is a need for methods that could leverage scenes with partial and noisy annotations.
%
\emph{Finally}, the popularity of commodity RGBD cameras has significantly simplified the acquisition of indoor scenes. This emerging scanning technique opens space for new applications such as online scene analysis with high fidelity scanning and reconstruction.  Availability of image data that come with RGBD scans also enables enhancing geometric representations with appearance information.


\section{Exploration and Organization}
\label{sec:exploration}
The rapidly growing number and diversity of digital 3D models in large online collections (e.g., TurboSquid, Trimble 3D Warehouse, etc.) have caused an emerging need to develop algorithms and techniques that effectively organize these large collections and allow users to interactively explore them.
For example, an architect can furnish a digital building by searching in databases organized according to furniture types, regions of interest and design styles, or an industrial designer can explore shape variations among existing products, when creating a new object.
Most existing repositories only support text-based search, relying on user-entered tags and titles. This approach suffers from inaccurate and ambiguous tags, often entered in different languages. While it is possible to try using shape analysis to infer consistent tags as discussed in Section \ref{sec:classification}, it is sometimes hard to convey stylistic and geometric variations using only text. An alternative approach is to perform shape-, sketch-, or image-based queries, however, to formulate such search queries the user needs to have a clear mental model of the shape that should be retrieved.
Thus, some researchers focus on providing tools for \emph{exploring} shape collections. Unlike search, exploration techniques do not assume a-priori knowledge of the repository content, and help the user to understand geometric, topological, and semantic variations within the collection.


\paragraph*{Problem statement and method categorization.}
Data exploration and organization is a classical problem in data analysis and visualization~\cite{Paulovich:2011:PLP}. Given a data collection, the research focuses on \emph{grouping and relating data points, learning the data variations in the collection, and organizing the collection into a structured form},
to facilitate retrieval, browsing, summarization, and visualization of the data, based on some efficient \emph{interfaces or metaphor}.

The first step to organizing model collections is to devise appropriate metrics to relate different data points. Various similarity metrics have been proposed in the past to relate entire shapes as well as local regions on shapes. In particular, previous sections of this document cover algorithms for computing global shape similarities (Section \ref{sec:classification}), part-wise correspondences (Section \ref{sec:segmentation}), and point-wise correspondences (Section \ref{sec:matching}). In this section, we will focus on techniques that take advantage of these correlations to provide different interfaces for exploring and understanding geometric variability in collections of 3D shapes.
We categorize the existing exploration approaches based on four aspects:
\begin{itemize}
\item \textbf{Metaphor:} a user interface for exploring shape variations. We will discuss five basic exploration interfaces, ones that use proxy shapes (templates), regions of interest, probability plots, query shapes, or continuous attributes.
\item \textbf{Shape comparison:} techniques used to relate different shapes. We will discuss techniques that use global shape similarities, and part or point correspondences.
\item \textbf{Variability:} shape variations captured by the system. Most methods we will discuss rely on geometric variability of shapes or parts. Some techniques also take advantage of topological variability, that is variance in number of parts or how they are connected (or variance in numbers of objects and their arrangements in scenes).
\item \textbf{Organization form:} a method to group shapes. We will discuss methods that group similar shapes to facilitate exploring intra-group similarities and inter-group variations, typically including clustering and hierarchical clustering.
\end{itemize}
Table~\ref{tab:exploration} summarizes several representative works in terms of these  aspects.
In the remaining part of this section we list several recent techniques grouped based on the exploration metaphor.

\begin{table}[!t]
\centering
\begin{tabular}{l|c|c|c|c} \hline
                         Method                & Meta. & Comp. & Var.    & Org.
 \\ \hline\hline
             \cite{Ovsjanikov:2011:ECV} & temp. & simi. & geom.  & n/a
 \\ \hline   \cite{Kim:2013:lpt}         & temp. & part & both    & cluster
 \\ \hline   \cite{Averkiou:2014:spm}  	   & temp. & part & both   & cluster
 \\ \hline   \cite{Kim:2012:FC}         & ROI & point & both      & n/a
 \\ \hline   \cite{Rustamov:2013:SD}  & ROI   & point & geom.     & n/a
 \\ \hline   \cite{Huang:2014:FMN}       & ROI & point & both      & cluster
 \\ \hline   \cite{Xu:2014:OHSC}         & ROI & simi. & topo.     & cluster
 \\ \hline   \cite{Fish:2014:MR}         & plot & part  & geom.     & cluster
 \\ \hline   \cite{Huang:2013:QOC}       & query & simi. & both   & hierarchy
 \\ \hline
\end{tabular}
\caption{A summary of several recent works over four aspects.
         \underline{Meta}phor: \underline{temp}lates, surface painted \underline{ROI}s,
         probability distribution \underline{plot}s, or \underline{query} shapes.
         Shape \underline{Comp}arison: shape \underline{simi}larity, \underline{part} or \underline{point} correspondence.
         \underline{Var}iability: \underline{geom}etry, \underline{topo}logy or \underline{both}.
         \underline{Org}anization Form: \underline{cluster} or \underline{hierarchy}.
         }
\label{tab:exploration}
\end{table}

\paragraph*{Template-based exploration.}
Component-wise variability in positions and scales of parts reveals useful information about a model collection. Several techniques use box-like templates to show variations among models of the same class. Ovsjanikov et al.~\cite{Ovsjanikov:2011:ECV} describe a technique for learning these part-wise variations without solving the challenging problem of consistent segmentation. First, they use a segmentation of a single shape to construct the initial template. This is the only step that needs to be verified and potentially fixed by the user. The next goal is to automatically infer deformations of the template that would capture the most important geometric variations of the models the collection.  They hypothesize that all shapes can be projected on a low-dimensional manifold based on their global shape descriptors. Finally, they reveal the manifold structure by deforming a template to fit to the sample points. Directions for interesting variations are depicted by arrows on the template and the shapes that correspond to current template configuration are presented to the user.

Descriptor-based approach described above assumes that all shapes share same parts and there exists a low-dimensional manifold that can be captured by deforming a single template. These assumptions do not hold for large and diverse collections of 3D models.  To tackle this challenge, Kim et al.~\cite{Kim:2013:lpt} proposed an algorithm for learning several part-based templates capturing multi-modal variability in collections of shapes. They start with an initial template that includes a super-set of all parts that might occur in a dataset, and jointly learn part segmentations, point-to-point surface correspondence and a compact deformation model. The output is \emph{a set of templates} that groups the input models into clusters capturing their styles and variations.

\begin{figure}[t]
\centering
    \includegraphics[width=1.0\columnwidth]{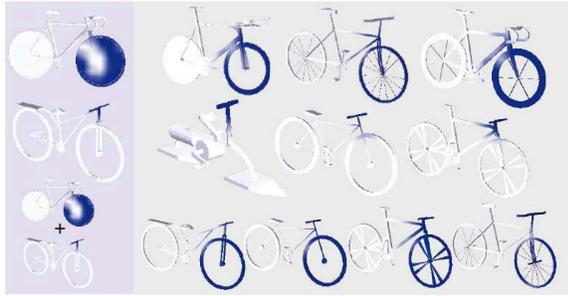}
    \vspace{-0.4cm}
    \caption{Shape exploration based on fuzzy correspondence. The user paints a region of interest (ROI) on a query shape (left column), and the method
    sorts models based on their similarity within the region (right).}
    \vspace{-0.6cm}
    \label{fig:kim_sig12_fc}
\end{figure}

\paragraph*{ROI-based exploration.}
Not all interesting variations occur at the scale of parts: they can occur at sub-part scale, or span multiple sub-regions from multiple parts. In these cases the user may prefer to select an arbitrary region on a 3D model and look for more models sharing similar regions of interest. Such detailed and flexible queries require a finer understanding of correspondences between different shapes. Kim et al.~\cite{Kim:2012:FC} propose fuzzy point correspondences to encode the inherent ambiguity in relating diverse shapes. Fuzzy point correspondences are represented by real values specified for all pair of points, indicating how well the points correspond.  They leverage transitivity in correspondence relationships to compute this representation from a sparse set of pairwise point correspondences.  The interface proposed by Kim et al. allows painting regions of interest directly on a surface, and the system retrieves similar regions or shows geometric variations in the selected region (see Figure~\ref{fig:kim_sig12_fc}).

One limitation of correspondence-based techniques is that they typically do not consider the entire collection when estimating shape differences.  Rustamov et al.~\cite{Rustamov:2013:SD} focus on a fundamental intrinsic representation for shape differences.  Starting with a functional map between two shapes, that is a map that describes change of functional basis, they derive a shape difference operator revealing detailed information about location, type, and magnitude of distortion induced by a map. This makes shape difference a quantifiable object that can be co-analyzed within a context of the entire collection. They show that this deeper understanding of shape differences can help in exploration. For example, one can embed shapes in a low-dimensional space based on shape differences, or use shape difference to interpolate variations by showing ``intermediate" shapes between two regions of interest.
To extend these technique to man-made objects, Huang et al.~\cite{Huang:2014:FMN} construct consistent functional basis for shape collections that exhibit large geometric and topological variability. They show that resulting consistent maps can capture discrete topological variability, such as variance in number of bars in the back of a chair.

\begin{figure}[t]
\centering
    \includegraphics[width=0.99\linewidth]{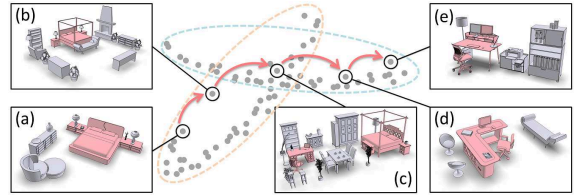}
    \caption{
    Focal-based scene clustering produces overlapping clusters, which is due to hybrid scenes possessing multiple focal points.
    An exploratory path, from (a) to (e), through the overlap, smoothly transit between the two scene clusters,
    representing bedroom and offices, respectively.}
    \vspace{-0.3cm}
    \label{fig:xu_sig14_expl}
\end{figure}

\paragraph*{ROI-based scene exploration.}
Recent works on organizing and exploring 3D visual data mostly focus on object collections. Exploring 3D scenes poses additional challenges since they typically exhibit more variance in structure. Unlike man-made objects that usually contain of a handful of object parts, scene usually includes tens to hundreds of objects, and most objects do not typically have a prescribed rigid arrangement. Thus, global scene similarity metrics, such as a graph kernel based technique by \cite{Fisher:2012:CSR} are limited to organizing datasets based on very high-level features, such as scene type.
Xu et al.~\cite{Xu:2014:OHSC} advocate that 3D scenes should be compared from a \emph{perspective} of a particular focal point which is a representative substructure of a specific scene category. Focal points are detected through contextual analysis of a collection of scenes, resulting in a clustering
of the scene collection where each cluster is characterized by its representative focal points (see Section~\ref{sec:scene}).
Consequently, the focal points extracted from a scene collection can be used to organize collection into an interlinked and well-connected cluster
formation, which facilitates scene exploration. Figure~\ref{fig:xu_sig14_expl} shows an illustration of such cluster-based organization
and an exploratory path transiting between two scene clusters/categories.

\paragraph*{Plot-based exploration.}
All aforementioned exploration techniques typically do not visualize the probabilistic nature of shape variations.  Fish et al.~\cite{Fish:2014:MR} study the configurations of shape parts from a probabilistic perspective, trying to indicate which shape variations are more likely to occur.  To learn the distributions of part arrangements, all shapes in the family are pre-segmented consistently.
The resulting set of probabilistic density functions (PDF) characterize the variability of relations and arrangements across different parts. A peak in a PDF curve represents a configuration of the related parts frequently appeared among several shapes in the family. The multiple PDFs can be used as interfaces to interactively explore the shape family from various perspectives.
Averkiou et al.~\cite{Averkiou:2014:spm}, use part structure inferred by this method to produce a low-dimensional part-aware embedding of all models. The user can explore interesting variations in part arrangements simply by moving the mouse over the 2D embedding. In addition, their technique allowed to synthesize novel shapes by clicking on empty spaces in the embedded space. At click the system would deform parts from neighboring shapes to synthesize a novel part arrangement.

\begin{figure}[t]
\centering
    \includegraphics[width=1.0\columnwidth]{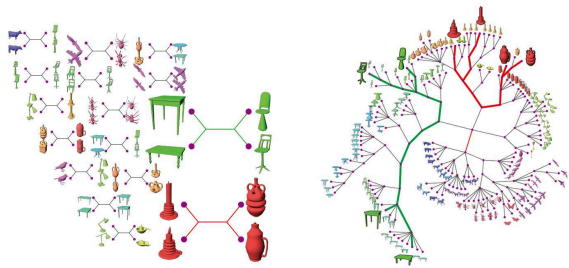}
    \vspace{-0.4cm}
    \caption{Given a set of heterogeneous shapes, a reliable qualitative similarity is derived from quartets composed of two pairs of objects (left).
    Aggregating such qualitative information from many quartets computed across the whole set leads to a categorization tree as a hierarchical
    organization of the input shape collection (right).}
    \vspace{-0.5cm}
    \label{fig:huang_sig12_quartet}
\end{figure}

\paragraph*{Query-based exploration.}
For a heterogeneous shape collection encompassing diverse object classes, it is typically not possible to capture shape part structure and correspondence. Even global shape similarity is not a very reliable feature, which makes organizing and exploring heterogeneous collections especially difficult.
To address this challenge, Huang et al.~\cite{Huang:2013:QOC} introduce qualitative analysis from the bioinformatics field. Instead of relying on quantitative distances, which may be unreliable between dissimilar shapes, the method considers more reliable \emph{qualitative similarity} derived from \emph{quartets} composed of two pairs of objects. The shapes that are paired in the quartet are close to each other and far from the shapes in the other pair, where distances are estimated from multiple shape descriptors.  They aggregate this topological information from many quartets computed across the entire shape collection, and construct a hierarchical \emph{categorization tree} (see Figure~\ref{fig:huang_sig12_quartet}). Analogous to the phylogenetic trees of species, the categorization tree of a shape collection provides an overview of the shapes about their mutual distance and hierarchical relations. Based on such organization, they also define the degree of separation chart for every shape in the collection and apply it for interactive shapes exploration.

\paragraph*{Attribute-based exploration.}
An alternative approach is to allow users interactively explore shapes with continuously valued semantic attributes.  Blanz and Vetter \cite{Blanz:1999:MMS} provide an interface to explore faces based on continuous facial attributes, such as ``smile'' or ``frown'', built upon the face parametric model (Section \ref{sec:recon}). Similarly, Allen et al.~\cite{Allen:2003:SHB} allow users explore the range of human bodies with features, such as height, weight, and age. Chaudhuri et al.'s~\cite{Chaudhuri:2013:ACC} interface enables exploration of shape parts according to learned strengths of semantic attributes (Figure \ref{fig:attribit}).

\section{Conclusion}
\label{sec:conclusion}

In this survey, we discussed the state-of-the-art on data-driven methods for 3D shape analysis and processing. We also presented the main concepts and methodologies used to develop such methods. We hope that this survey will act as a tutorial that will help researchers develop new data-driven algorithms related to shape processing. There are several exciting research directions that have not been sufficiently explored so far in our community that we discuss below:

\paragraph*{Joint analysis of 2D and 3D data.}
Generating 3D content from images requires building mappings from 2D to 3D space. The problem is largely ill-posed, however, with the help of the vast amount of 2D images available on the web, effective priors can be developed to map 2D visual elements or features to 3D shape and scene representations. Initial attempts to build alignments between 2D and 3D data are the recent works by Su et al \cite{Su:2014:EID} and Aubry et al.\cite{Aubry14}, which can further inspire more work on this topic. Another possibility is to jointly analyze shape and texture data. The work of co-segmenting textured 3D shapes by Yumer et al. \cite{Yumer:CST:2014} is one such example.
 Following this line, it would be interesting to jointly analyze and process multi-modal visual data, including depth scans and videos. The key challenges is how to integrate the heterogeneous information in a unified learning framework.

\paragraph*{Better and scalable shape analysis techniques.} Many data-driven applications rely on high-quality shape analysis results, particularly in segmentations and correspondences. We believe it is important to further advance the research in both directions. This includes designing shape analysis techniques for specific data and/or making them scalable to gigantic datasets.

\paragraph*{From geometry to semantics and vice versa.}
Several data-driven methods have tried to map 2D and 3D geometric data to high-level concepts, such as shape categories, semantic attributes, or part labels. Existing methods deal with cases where only a handful of different entities are predicted for input shapes or scenes. Scaling these methods to handle thousands or more categories, part labels and other such entities, as well as approaching human performance is an open problem. The opposite direction is also interesting and insufficiently explored: generating or editing shapes and scenes based on high-level specifications, such as shape styles, attributes, or even natural language, potentially combined with other input, such as sketches and interactive handles. WordsEye \cite{Coyne:2001:WAT} was an early attempt to bridge this gap, yet requires largely manual mappings. The more recent work by \cite{Chaudhuri:2013:ACC} handles only shape part replacements driven by linguistic attributes.

\paragraph*{Understanding function from geometry.}
The geometry of a shape is strongly related to its functionality including its relationship to human activity. Thus, analyzing shapes and scenes requires some understanding of their function. The recent work by Laga et al. \cite{Laga:2013:GCS} and Kim et al. \cite{Kim14} are important examples of data-driven approaches that take into account functional aspects in shape analysis. In addition, data-driven methods can guide the synthesis of shapes that can be manufactured or 3D printed based on given functional specifications; an example of such attempt is the work by Schulz et al \cite{Schulz:2014:DFE}.

\paragraph*{Data-driven shape abstractions.}
It is relatively easy for humans to communicate the essence of shapes with a few lines, sketches, and abstract forms. Developing methods that can build such abstractions automatically has significant applications in shape and scene visualization, artistic rendering, and shape analysis. There are a few data-driven approaches to line drawing \cite{Cole:2008:WDP,Kalogerakis:2009:DDC,Kalogerakis:2012:LHP}, saliency analysis \cite{Chen:2012:SPO}, surface abstraction \cite{Yumer:2012:CSC}, and viewpoint preferences \cite{Secord:2011:PMO} related to this goal. Matching the human performance in these tasks is still a largely open question, while synthesizing and editing shapes using shape abstractions as input remains a significant challenge.

\paragraph*{Feature learning.}
Several shape and scene processing tasks depend on designing geometric descriptors for points and shapes, as we show in Section~\ref{fig:overview_seg}. In general, it seems that some descriptors work well in some specific classes, but fail in several others. A main issue is that there are no geometric features that can serve as reliable mid- or high-level representations of shapes.
Recent work in computer vision shows that features can be learned from data in the case of 2D and 3D images \cite{Yu:FLI:2010,Lai:2013:UFL}, thus a promising direction is to extend this work for learning feature representations from raw 3D geometric data.

\begin{table*}[t!]
\scriptsize
  \centering
    \begin{tabular*}{\textwidth}{l|c|c|c|c|c|c|c|c|c}
    \hline
    \multirow{2}{*}{Work} & \multicolumn{3}{c|}{Training data} & \multicolumn{2}{c|}{Feature} & \multirow{2}{*}{Learning model/approach}& \multirow{2}{*}{Learning type} & \multirow{2}{*}{Learning outcome} & \multirow{2}{*}{Application}\\
           \cline{2-6}
                          & Rep. & Preproc. & Scale & Type & Sel. & \multicolumn{1}{r|}{} & \multicolumn{1}{r|}{} & \multicolumn{1}{r|}{} \\
    \hline
    \hline
    \cite{Funkhouser:2005:SBR}  & Point     & No            & Thousands & Local     & No & SVM classifier                   & Supervised        & Object classifier & Classification   \\ \hline
    \cite{Bronstein:2011:SGGW}  & Mesh      & No            & Thousands & Local     & No & Similarity Sensitive Hashing     & Supervised        & Distance metric   & Classification   \\ \hline
    \cite{Huang:2013:FSL}       & Mesh      & Pre-align.	& Thousands	& Local     & No & Max-marginal distance learning   & Semi-supervised	& Distance metric	& Classification   \\ \hline
    \cite{Kalogerakis:2010:LMS} &Mesh	&No	&Tens	&Local	&Yes	&Jointboost classifier	&Supervised	&Face classifier	&Segmentation \\ \hline
    \cite{van-Kaick:2011:PKC}   &Mesh	&Yes	&Tens	&Local	&Yes	&Gentleboost classifier	&Supervised	&Face classifier	&Segmentation \\ \hline
    \cite{Benhabiles:2011:LBE}	&Mesh	&No	&Tens	&L.\&G.	&Yes	&Adaboost classifier	&Supervised	&Boundary classifier	&Segmentation \\ \hline
    \cite{Zhige:2014:SSL}	&Mesh	&No	&Hundreds	&Local	&Yes	&Feedforward neural networks	&Supervised	&Face/patch classifier	&Segmentation \\ \hline
    \cite{Xu:2014:TSS}	&Mesh	&Pre-seg.	&Tens	&Local	&No	&Sparse model selection	&Supervised	&Segment similarity	&Segmentation \\ \hline
    \cite{Lv:2012:SMS}	&Mesh	&No	&Tens	&Local	&Yes	&Entropy regularization	&Semi-supervised	&Face classifier	&Segmentation \\ \hline
    \cite{Wang:2012:ACS}	&Mesh	&Pre-seg.	&Hundreds	&Local	&No	&Active learning	&Semi-supervised	&Segment classifier	&Segmentation \\ \hline
    \cite{Wang:2013:PAS} &Image	&Labeled parts	&Hundreds	&Local	&No	&2D shape matching	&Supervised	&2D shape similarity	&Segmentation \\ \hline
    \cite{Hu:2012:CSS}	&Mesh	&Over-seg.	&Tens	&Local	&Yes	&Subspace clustering	&Unsupervised	&Patch similarity	&Seg. / Corr. \\ \hline
    \cite{Sidi:2011:CS}	&Mesh	&Pre-seg.	&Tens	&Local	&No	&Spectral clustering	&Unsupervised	&Seg. simi./classifier	&Seg. / Corr. \\ \hline
    \cite{Xu:2010:SCS}	&Mesh	&Part	&Tens	&Struct.	&No	&Spectral clustering	&Unsupervised	&Part proportion simi.	&Seg. / Corr. \\ \hline
    \cite{van-Kaick:2013:CHA}	&Mesh	&Part	&Tens	&Struct.	&No	&Multi-instance clustering	&Unsupervised	&Seg. hier. simi.	&Seg. / Corr. \\ \hline
    \cite{Golovinskiy:2009:CS}	&Mesh	&No	&Tens	&Global	&No	&Global shape alignment	&Unsupervised	&Face similarity	&Seg. / Corr. \\ \hline
    \cite{Huang:2011:JSS}	&Mesh	&Pre-seg.	&Tens	&Local	&No	&Joint part matching	&Unsupervised	&Segment similarity	&Seg. / Corr. \\ \hline
    \cite{Huang:2014:FMN}	&Mesh	&Init. corr.	&Tens	&Global	&No	&Consistent func. map networks	&Unsupervised	&Segment similarity	&Seg. / Corr. \\ \hline
    \cite{Kim:2013:lpt}	&Mesh	&Template	&Thousands	&Local	&No	&Shape alignment	&Semi-supervised	&Templates	&Seg. / Corr. \\ \hline
    \cite{MATTAUSCH:2014:ODC}	&Mesh	&Over-seg.	&Hundreds	&Local	&No	&Density-based clustering	&Unsupervised	&Patch similarity	&Recognition \\ \hline
    \cite{Nguyen:2011:CSM}   &Mesh	&Init. corr.	&Tens	&L.\&G.	&No	&Inconsistent map detection	&Unsupervised	&Point similarity	&Corr. / Expl. \\ \hline
    \cite{Huang:2012:OAE}	&Mesh	&Init. corr.	&Tens	&L.\&G.	&No	&MRF joint matching	&Unsupervised	&Point similarity	&Corr. / Expl. \\ \hline
    \cite{Kim:2012:FC}	&Mesh	&Pre-align.	&Tens	&Global	&No	&Spectral matrix recovery	&Unsupervised	&Point similarity	&Corr. / Expl. \\ \hline
    \cite{Huang:2013:SDP}	&Mesh	&Init. corr.	&Tens	&Global	&No	&Low-rank matrix recovery	&Unsupervised	&Point similarity	&Corr. / Expl. \\ \hline
    \cite{Ovsjanikov:2011:ECV}	&Mesh	&Part	&Hundreds	&Global	&No	&Manifold learning	&Unsupervised	&Parametric model	&Exploration \\ \hline
    \cite{Rustamov:2013:SD}	&Mesh	&Map	&Tens	&None	&N/A	&Functional map analysis	&Unsupervised	&Difference operator	&Exploration \\ \hline
    \cite{Fish:2014:MR}	&Mesh	&Labeled parts	&Hundreds	&Struct.	&No	&Kernel Density Estimation	&Supervised	&Prob. distributions	&Expl. / Synth. \\ \hline
    \cite{Averkiou:2014:spm}	&Mesh	&\cite{Kim:2013:lpt}	&Thousands	&Struct.	&No	&Manifold learning	&Unsupervised	&Parametric models	&Expl. / Synth. \\ \hline
    \cite{Huang:2013:QOC}	&Mesh	&No	&Hundreds	&Global	&No	&Quartet analysis and clustering	&Unsupervised	&Distance measure	&Organization \\ \hline
    \cite{Blanz:1999:MMS}	&Mesh	&Pre-align.	&Hundreds	&Local	&No	&Principal Component Analysis	&Unsupervised	&Parametric model	&Recon. / Expl. \\ \hline
    \cite{Allen:2003:SHB}	&Point	&Pre-align.	&Hundreds	&Local	&No	&Principal Component Analysis	&Unsupervised	&Parametric model	&Recon. / Expl. \\ \hline
    \cite{Hasler:2009:SSR}	&Point	&Pre-align.	&Hundreds	&Local	&No	&PCA \& linear regression	&Unsupervised	&Parametric model	&Recon. / Expl. \\ \hline
    \cite{Pauly:2005:ESC}	&Mesh	&Pre-align.	&Hundreds	&Global	&No	&Global shape alignment	&Unsupervised	&Shape similarity	&Reconstruction \\ \hline
    \cite{Nan:2012:SAC}	&Point	&Labeled parts	&Hundreds	&Struct.	&No	&Random Forest Classifier	&Supervised	&Object classifier	&Reconstruction \\ \hline
    \cite{Shen:2012:SRP}	&Mesh	&Labeled parts	&Tens	&Global	&No	&Part matching	&Unsupervised	&Part detector	&Reconstruction \\ \hline
    \cite{Kim:2012:AIE}	&Point	&Labeled parts	&Tens	&Local	&No	&Joint part fitting and matching	&Unsupervised	&Object detector	&Reconstruction \\ \hline
    \cite{Salas-Moreno:2013:SLAM}	&Mesh	&No	&Tens	&L.\&G.	&No	&Shape matching	&Unsupervised	&Object detector	&Reconstruction \\ \hline
    \cite{Xu:2011:PMO}	&Mesh	&Labeled parts	&Tens	&Struct.	&No	&Structural shape matching	&Unsupervised	&Part detector	&Modeling \\ \hline
    \cite{Aubry14}	&Mesh	&Projected	&Thousands	&Visual	&No	&Linear Discriminant Analysis	&Supervised	&Object detector	&Recognition \\ \hline
    \cite{Su:2014:EID}	&Mesh	&Projected	&Tens	&Visual	&No	&Shape matching	&Unsupervised	&2D-3D correlation	&Reconstruction \\ \hline
	\cite{Chaudhuri:2010:DDS} &Mesh	&No	&Thousands	&Global	&No	&Shape matching	&Unsupervised	&Part detector	&Modeling \\ \hline
	\cite{Chaudhuri:2011:prabm}	&Mesh	&\cite{Kalogerakis:2010:LMS}	&Hundreds	&Local	&No	&Bayesian Network	&Unsupervised	&Part reasoning model	&Modeling \\ \hline
	\cite{Xie:2013:S2D}	&Mesh	&Labeled parts	&Tens	&Struct.	&No	&Contextual part matching	&Unsupervised	&Part detector	&Modeling \\ \hline
	\cite{Kalogerakis:2012:PMC}	&Mesh	&\cite{Kalogerakis:2010:LMS}	&Hundreds	&L.\&G.	&No	&Bayesian Network	&Unsupervised	&Shape reasoning model	&Synthesis \\ \hline
	\cite{Xu:FDS:2012}	&Mesh	&Part	&Tens	&Struct.	&No	&Part matching	&Unsupervised	&Part similarity	&Synthesis \\ \hline
	\cite{Talton:2012:LDP}	&Mesh	&Labeled parts	&Tens	&Struct.	&No	&Structured concept learning	&Unsupervised	&Probabilistic grammar	&Synthesis \\ \hline
	\cite{Yumer:2012:CSC}	&Mesh	&No	&Tens	&Global	&No	&Shape matching	&Unsupervised	&Shape abs. similarity	&Modeling \\ \hline
	\cite{Yumer:2014:CCH}	&Mesh	&Pre-seg.	&Tens	&Local	&No	&Segment matching	&Unsupervised	&Segment abs. simi.	&Modeling \\ \hline
    \cite{Chaudhuri:2013:ACC}   &Mesh	&\cite{Kalogerakis:2010:LMS}	&Hundreds	&L.\&G.	&No	&SVM ranking        &Supervised	        & Ranking metric	& Model. / Expl. \\ \hline
	\cite{Fisher:2011:CSR}	&Scene	&Labeled obj.	&Tens	&Struct.	&No	&Relevance feedback	&Supervised	&Contextual obj. simi.	&Classification \\ \hline
    \cite{Fisher:2012:CSR}	&Scene	&Labeled obj.	&Hundreds	&Struct.	&No	&Bayesian Network	&Supervised	&Mixture models	&Synthesis \\ \hline
    \cite{Xu:2013:S2S}	&Scene	&Labeled obj.	&Hundreds	&Struct.	&No	&Frequent subgraph mining	&Unsupervised	&Frequent obj. groups	&Modeling \\ \hline
    \cite{Xu:2014:OHSC}	&Scene	&Labeled obj.	&Hundreds	&Struct.	&No	&Weighted subgraph mining	&Unsupervised	&Distinct obj. groups	&Org. / Expl. \\ \hline
    \cite{Liu:2014:CCS}	&Scene	&Labeled hier.	&Tens	&Struct.	&No	&Probabilistic learning	&Supervised	&Probabilistic grammar	&Seg. / Corr. \\
    \hline
    \end{tabular*}%
  \caption{Comparison of various works on data-driven shape analysis and processing.
		   For each work, we summarize over the criterion set defined for data-driven methods: training data (encompassing data representation, preprocessing and scale),
           feature (including feature type and whether feature selection is involved), learning model or approach, learning type (supervised, semi-supervised, and unsupervised),
           learning outcome (e.g., a classifier or a distance metric), as well as its typical application scenario. See the text for detailed explanation of the criteria.
           Some works employ another work as a pre-processing stage (e.g.,~\cite{Chaudhuri:2013:ACC} requires the labeled segmentation produced by~\cite{Kalogerakis:2010:LMS}).
           There are four types of features including local geometric features (Local), global shape descriptors (Global), both local and global shape features (L.\&G.),
           structural features (Struct.) as well as 2D visual features (Visual).}
  \label{tab:compare}%
\end{table*}

\section*{Biographical sketches}
\label{sec:bios}

\paragraph*{Kai Xu} received his PhD in Computer Science at National University of Defense Technology (NUDT). He is currently a postdoctoral researcher at Shenzhen Institutes of Advanced Technology and also holds a faculty position at NUDT.
During 2009 and 2010, he visited Simon Fraser University, supported by the Chinese government. His research interests include geometry processing and geometric modeling, especially on methods that utilize large collections of 3D shapes.
He served on program committees for SGP, PG and GMP.

\vspace{-.3cm}

\paragraph*{Vladimir G. Kim} received his PhD in the Computer Science Department at Princeton University and is currently a postdoctoral scholar at Stanford University. His research interests include geometry processing and analysis of shapes and collections of 3D models. He received his B.A. degree in Mathematics and Computer Science from Simon Fraser University in 2008. Vladimir is a recipient of the Siebel Scholarship and the NSERC Postgraduate Scholarship. He was also on the International Program Committee for SGP 2013 and SGP 2014.

\vspace{-.3cm}

\paragraph*{Qixing Huang} is a research assistant professor at TTI Chicago. He earned his PhD from the Department of Computer Science at Stanford University in 2012. He obtained both MS and BS degrees in Computer Science from Tsinghua University in 2005 and 2002, respectively. His research interests include data-driven geometry processing and co-analysis of shapes and collections of 3D models using convex optimization techniques. He was a winner of the Best Paper Award from SGP 2013 and the Most Cited Paper Award for the journal Computer-Aided Geometric Design in 2011 and 2012. He served on program committees for SGP, PG and GMP.

\vspace{-.3cm}

\paragraph*{Evangelos Kalogerakis} is an assistant professor in computer science at the University of Massachusetts Amherst. His research deals with automated analysis and synthesis of 3D visual content, with particular emphasis on machine learning techniques that learn to perform these tasks by combining data, probabilistic models, and prior knowledge. He obtained his PhD from the University of Toronto in 2010 and BEng from the Technical University of Crete in 2005. He was a postdoctoral researcher at Stanford University from 2010 to 2012.  He served on program committees for EG 2014 and 2015, SGP 2012, 2014 and 2015. His research is supported by NSF (CHS-1422441).

\vspace{-.3cm} 


\bibliographystyle{eg-alpha}

\bibliography{references_intro,references_matching,references_reconstruction,references_scene_expl,references_segmentation,references_modeling,references_search}


\end{document}